\documentclass[
    10pt,%
    aps,%
    pre,%
    twocolumn,%
    groupedaddress,%
    showpacs,%
    superscriptaddress,%
    amssymb,amsmath,%
    noeprint,%
]{revtex4-2}

\usepackage{graphicx}

\usepackage{hyperref}
 \hypersetup{
     colorlinks=true,
     linkcolor=blue,
     urlcolor=blue,
     citecolor=blue
 }
 
\usepackage{cleveref}

\newcommand{\rme}{{\rm e}}
\newcommand{\rmc}{{\rm c}}
\newcommand{\rmd}{{\rm d}}
\newcommand{\rmi}{{\rm i}}
\newcommand{\rms}{{\rm s}}
\newcommand{\op}[1]{| {#1} \rangle \langle {#1}| }

\begin{document}

\title{Chaos from quantum bath fluctuations}

\author{Ilan Baud}
\affiliation{Institute of Physics, \'{E}cole Polytechnique F\'{e}d\'{e}rale de Lausanne (EPFL), 1015 Lausanne, Switzerland}

\author{Tamoghna Ray}
\affiliation{International Centre for Theoretical Sciences, Tata Institute of Fundamental Research, Bangalore 560089, India}

\author{Mahaveer Prasad}
\affiliation{Science, Mathematics and Technology Cluster, Singapore University of Technology and Design, 487372 Singapore}
\affiliation{Centre for Quantum Technologies, National University of Singapore 117543, Singapore}

\author{Manas Kulkarni}
\affiliation{International Centre for Theoretical Sciences, Tata Institute of Fundamental Research, Bangalore 560089, India}

\author{Camille Aron}
\email{aron@ens.fr} 
\affiliation{Laboratoire de Physique de l’\'{E}cole Normale Sup\'{e}rieure, ENS, Universit\'{e} PSL, \\ CNRS, Sorbonne Universit\'{e}, Universit\'{e} Paris Cit\'{e}, 75005 Paris, France}
\affiliation{Institute of Physics, \'{E}cole Polytechnique F\'{e}d\'{e}rale de Lausanne (EPFL), 1015 Lausanne, Switzerland}

\date{\today} 

\begin{abstract}
The effect of a large environment on a finite-size quantum mechanical system is two-fold: It brings dissipation, but also fluctuations of thermal and quantum origin. While dissipation tends to stabilize the dynamics, we question if and how environmental quantum fluctuations can generate chaos in an otherwise classically non-chaotic system. We work out a paradigmatic model of quantum optics: the dissipative Dicke model, where a large spin interacts with a dissipative harmonic mode. We dial in the classical/quantum correspondence by working in the semiclassical regime at large but finite spin. We demonstrate that, starting from a classically regular phase space in the superradiant regime, quantum noise can generate a strange attractor with fractal dimension and a positive Lyapunov exponent. We unveil the deep connection with shear-induced chaos that was recently developed in the mathematical community. 
\end{abstract}
    \maketitle

\paragraph*{Introduction.}
The impact of environments on signatures of quantum chaos has long been of considerable interest~\cite{Sa_2026}, yet remains only partially understood.
Dissipation is generally expected to suppress chaos, as identified early on in dissipative quantum maps~\cite{Dittrich_1987, Braun_2001}. More recently in interacting many-body systems, local dissipation was shown to continuously reduce the top Lyapunov exponent of the SYK model, eventually rendering the infinite-temperature dynamics non-chaotic at sufficiently large dissipation strength~\cite{Jacobus_2024}. At the same time, substantial progress has been made in identifying spectral signatures of dissipative quantum chaos~\cite{Prosen_2020, Sa_2023,PRXQuantum2023}.

Dissipative environments can also appear to favor chaotic behavior. In the quantum kicked rotor, a paradigmatic model for the interplay between quantum interference and classically chaotic dynamics, coupling to an environment destroys dynamical localization~\cite{SH95}. However, this merely reflects the suppression of quantum coherence on an already classically chaotic substrate rather than the generation of a new instability.

Beyond dissipation, environments also introduce fluctuations. This raises the question of whether quantum fluctuations originating from a zero-temperature bath can destabilize attracting fixed points and induce chaotic dynamics starting from an otherwise classically non-chaotic substrate.
On the spectral side, it has recently been shown that quantum jumps in Lindblad dynamics can promote random-matrix statistics~\cite{Gupta_2024}.
Here, we address the dynamical counterpart of this question in the context of the Dicke model~\cite{Dicke_1954,Gross_1982,Kirton_2019}, a paradigmatic spin–boson system in quantum optics realized with several experimental platforms~\cite{Baumann_2010,Jaako_2016,Baden_2014,Klinder_2015,Zhang_2017,Zhang_2018,Cohn_2018,Safavi_2018}.

The specificity of the closed Dicke model is that, although non-integrable at finite spin–boson coupling, studies of both its classical dynamics and the spectral statistics of its quantum Hamiltonian have revealed an unusual robustness of integrable signatures as the coupling is increased, with the onset of chaos delayed until the thermodynamic phase transition~\cite{Emary_PRE_2003,Emary_PRL_2003}. As such, the Dicke model provides a single-system realization of both the Berry–Tabor and Bohigas–Giannoni–Schmit (BGS) conjectures~\cite{BerryTabor_1977,BGS_1984}.
This robustness of integrable spectral statistics persists in the \emph{dissipative} Dicke model, where dissipation is included. Exact diagonalization of the complex Liouvillian spectrum at finite spin $s=5$ shows a delayed crossover from 2D Poisson to GinUE statistics at coupling strengths comparable to the dissipative phase transition~\cite{Prasad_2022}.
However, considering that the classical limit of the dissipative model has attractive fixed points and hence non-chaotic dynamics, Ref.~\cite{Villasenor_2024} notably challenged the validity of the Grobe–Haake–Sommers (GHS) conjecture, which posits, in analogy with the BGS conjecture for closed systems, that Liouvillian spectral correlations on the quantum side should correspond to chaotic dynamics in the classical limit~\cite{GHS_1988}. Subsequent studies~\cite{Villase_2025,Ferrari_2025,Santos_2026} have emphasized the distinction between transient and steady-state chaotic behavior.

In this Letter, we revisit the semiclassical regime of the dissipative Dicke model at large spin $s$. We demonstrate that $1/s$ quantum fluctuations induced by the environment, acting as a noise, can destabilize attractive fixed points into random attractors with positive Lyapunov exponents~\cite{eckmann1985ergodic,lyapunov1992general,arnold1986lyapunov,tel2006chaotic,lakshmanan2012nonlinear}. This raises the possibility that a GHS-like semiclassical-quantum correspondence could be formulated at finite $s$.
We interpret this mechanism in the framework of shear-induced chaos~\cite{Lin_2008}, whereby stable attractors become chaotic under time-dependent perturbations, a phenomenon rigorously established in a variety of deterministic and stochastic setting~\cite{Baxendale_1994,Ott_2008,Engel_2018,Engel_2019,Baxendale_2023,Engel_2023,Blumenthal_2023}.

We first introduce the dissipative Dicke model and the Wigner phase-space formulation of quantum mechanics that allows a controlled derivation of the semiclassical dynamics at large but finite spin $s$. This leads to the stochastic Dicke model, in which quantum bath fluctuations enter as stochastic noise terms. We then present numerical evidence for chaos at finite spin, characterizing the resulting random attractor through (i) pullback dynamics, (ii) fractal dimension, and (iii) Lyapunov exponents. Finally, we discuss these results in the context of shear-induced chaos and outline broader implications beyond the Dicke model, highlighting how this connection may offer a new perspective on the characterization of quantum chaotic dynamics.

\paragraph*{Dissipative Dicke model.}
The dissipative Dicke model describes the electric-dipole interaction between a collective spin formed by an ensemble of quantum emitters and a single leaky cavity mode.
The dynamics of its density matrix $\rho(t)$ are described by a Lindblad master equation $\partial_t \rho(t) = \mathcal{L}\,\rho(t)$ with the Liouvillian~\cite{breuer2002theory} (we set $\hbar = 1$)
\begin{align} \label{eq:Liouvillian}
      \mathcal{L}\, \star = -\rmi \,[H,\star] + \kappa \left(2 a \star a^\dagger -  \left\{a^\dagger a, \star \right\} \right),
\end{align}
and the Dicke Hamiltonian
\begin{align}
    H = \omega_\rmc a^\dagger a +
    \omega_\rms S_z +\lambda \sqrt{\frac{2}{s}}(a+a^\dagger)
    S_x,
    \label{eq:Dicke_hamiltonian}
\end{align}
where $a$ ($a^\dagger$) is a bosonic annihilation (creation) operator. 
The operators $S_i$ ($i=x,y,z$), satisfying $[S_i, S_j] = \rmi \epsilon_{ijk} S_k$, are the generators of the spin-$s$ irreducible representation of SU(2) of fixed dimension $2s+1$. 
$\omega_\rmc$ and $\omega_\rms$ are the bosonic and spin energy splittings, respectively,
and the spin-boson coupling $\lambda$ is multiplied by $1/\sqrt{s}$ to ensure a nontrivial thermodynamic limit.
The non-unitary part of the Liouvillian describes boson loss at a rate $\kappa>0$. This choice of dissipator is physically justified in driven cavity-QED realizations, where the environment dominantly interacts directly with the bare cavity mode through photon leakage. The resulting Lindblad equation can thus be regarded as an effective nonequilibrium description of a driven-dissipative cavity~\cite{Baumann_2010}.

The Dicke Hamiltonian is $\mathbb{Z}_2$-symmetric: it is invariant under $(a, S_x,S_y) \mapsto (-a, -S_x, -S_y)$. 
Note that these are the counter-rotating terms, $a^\dagger S^+$ and $a S^-$, that break the U(1) symmetry and the integrability of the model.
The Liouvillian inherits a weak $\mathbb{Z}_2$ symmetry which is spontaneously broken in the thermodynamic limit at the critical coupling $\lambda_\rmc^2 = \frac14 \frac{\omega_\rms}{\omega_\rmc} \left( \omega_\rmc^2 + \kappa^2 \right)$~\cite{Carmichael_2007,Keeling_2019,Roses_2020}.
This corresponds to a second-order dissipative phase transition between a normal phase for $\lambda < \lambda_\rmc$, where the steady-state boson expectation value vanishes $\langle a \rangle = 0$, and a superradiant phase for $\lambda > \lambda_\rmc$, where it acquires a finite expectation value $\langle a \rangle \neq 0$.

The initial conditions are mostly inconsequential to our discussion. Unless stated otherwise, we initialize the system in the product state of the bosonic vacuum and the spin at the south pole,  $\rho_0 = \op{0} \otimes\op{\downarrow}$.

\begin{figure}
    \centering
    \begin{minipage}{.49\linewidth}
     \includegraphics[width=\linewidth]{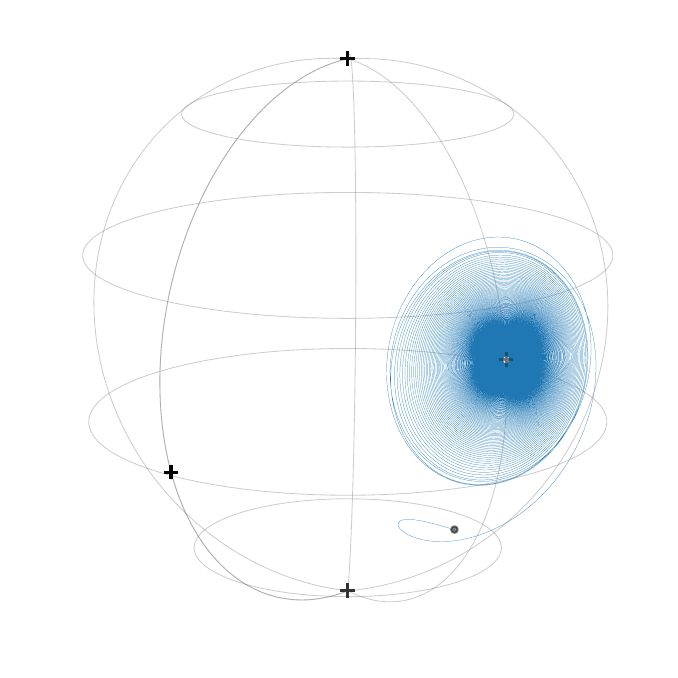}
        \vspace*{-45mm}
            \begin{center}
                (a) $s\to\infty$\hfill\textcolor{white}{.}
            \end{center}
        \vspace*{30mm}
    \end{minipage}
    \hfill
    \begin{minipage}{.49\linewidth}
          \includegraphics[width=\linewidth]{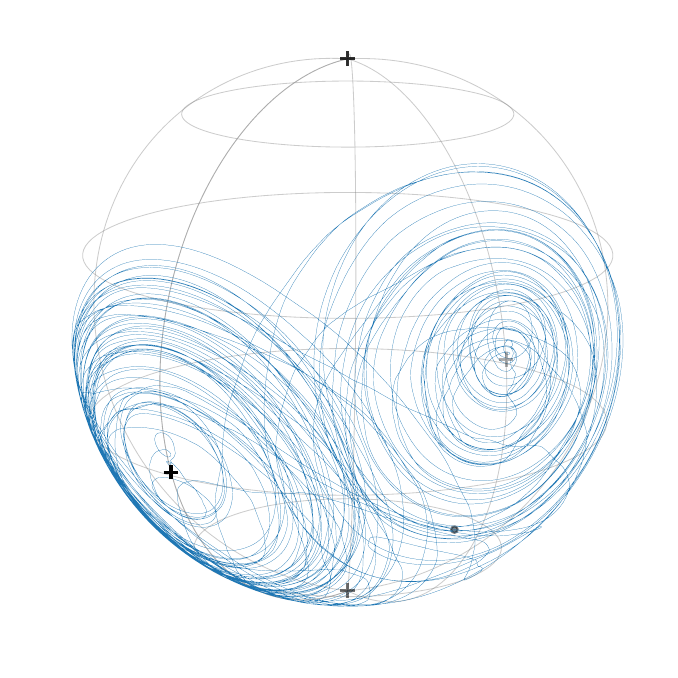}
        \vspace*{-45mm}
            \begin{center}
                (b) $s = 4$\hfill\textcolor{white}{.}
            \end{center}
        \vspace*{30mm}
    \end{minipage}
    \begin{minipage}[b]{.53\linewidth}
        \includegraphics[width=\linewidth]{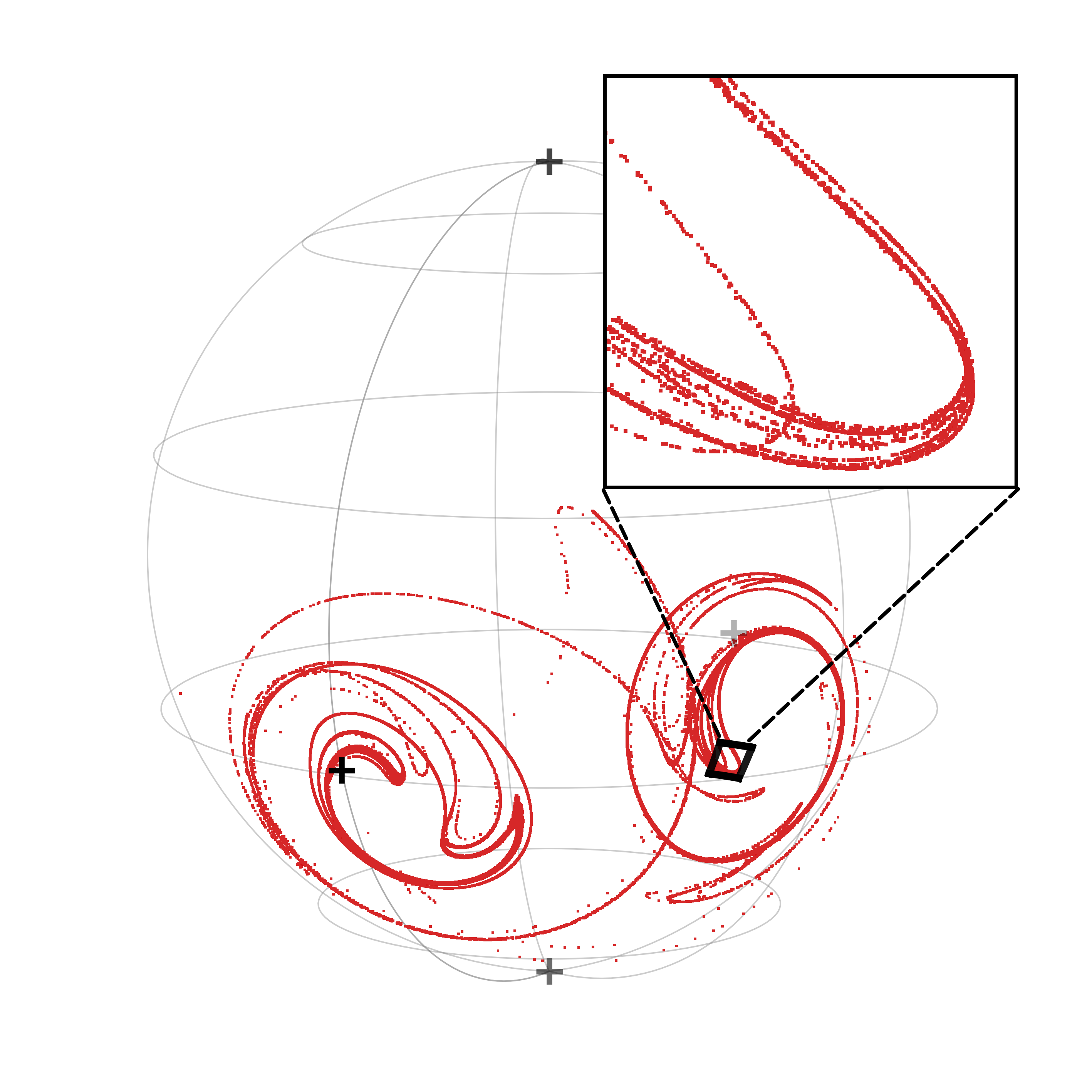}
      \vspace*{-45mm}
            \begin{center}
                (c) $s=4$ \hfill\textcolor{white}{.}
            \end{center}
        \vspace*{30mm}
    \end{minipage}
    \hfill
    \begin{minipage}[b]{.44\linewidth}
        \includegraphics[width=\linewidth]{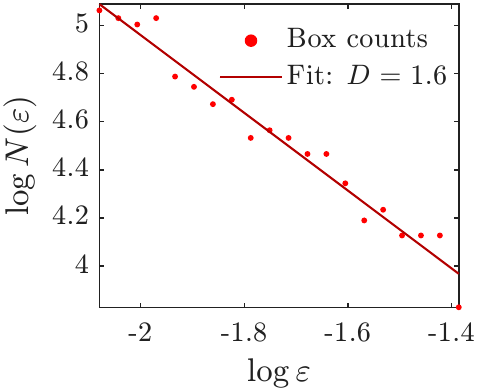}
        \vspace*{-40mm}
            \begin{center}
                (d)\hfill\textcolor{white}{.}
            \end{center}
        \vspace*{30mm}
    \end{minipage}
\caption{
Noise-induced chaotic attractor for $\lambda > \lambda_\rmc$.
Representative spin trajectories: (a) classical dissipative Dicke model in the limit $s \to \infty$, and (b) stochastic Dicke model at $s=4$, both initialized at the black dot and with dynamics governed by Eqs.~(\ref{eq:eom_x_p_sigma}). 
Black crosses denote the four classical fixed points.
(c) Random strange attractor at $s=4$ (inset: zoom), obtained from pullback dynamics for a fixed noise realization and $N_0 = 10^5$ initial conditions. 
(d) Box-counting estimate of the fractal dimension of the strange attractor shown in (c), embedded in a four-dimensional phase space, yielding $D \approx 1.6$.
Parameters: $\omega_\rms = \omega_\rmc =1$, $\kappa = 0.5$, $\lambda = 1.0$.
}
    \label{fig:attractor}
\end{figure}

\paragraph*{Stochastic Dicke model.}
We derive the semiclassical description of the dissipative Dicke model by means of the truncated Wigner approximation (TWA), which allows to systematically account for leading-order quantum fluctuations on top of the classical dynamics~\cite{Polkovnikov_2010}.
It relies on an exact phase-space representation of $\rho(t)$ in terms of the Wigner function $W(\alpha, \alpha^*,\vec \sigma;t)$, where $\alpha$, $\alpha^*$ are the complex amplitudes associated with the local coherent states of the boson and the spin variables collected in $\vec \sigma$ can be constructed from representing the spin in terms of two Schwinger bosons. 
The semiclassical equations associated with Eq.~(\ref{eq:Liouvillian}) are obtained by first rescaling $\alpha \to \sqrt{s} \alpha$ and $\vec \sigma \to s \vec \sigma$, and organizing the contributions in powers of $1/s$. The expansion terminates at order $1/s^2$. 
The TWA consists in neglecting those terms, that are associated with intrinsic quantum fluctuations of the dynamics, while retaining the $\kappa/s$ terms describing quantum fluctuations generated by the environment.
We refer the reader to Sec.~\ref{app:sec:TWA} of the Supplemental Material (SM) for the details of this construction.
At large $s$, the dynamics of the Wigner function are governed by a Fokker-Planck equation. The latter can be mapped to the following set of coupled stochastic differential equations (SDEs), defining what we refer to as the stochastic Dicke model. They describe the noisy nonequilibrium dynamics of a classical oscillator with coordinates $x := (\alpha^* + \alpha)/\sqrt{2}$, $p:=\rmi (\alpha^*-\alpha)/\sqrt{2}$ coupled to a classical top with coordinates $\vec \sigma := (\sigma_x,\sigma_y,\sigma_z)$, and read
\begin{align}\label{eq:eom_x_p_sigma}
\everymath{\displaystyle}
\left\{
\begin{array}{rl}
     \partial_t  x &=  \omega_\rmc \, p - \kappa \, x - \sqrt{{\kappa}/{s}} \, \xi_1\\
    \partial_t p &= - \kappa \, p -\omega_\rmc \, x  - 2 \lambda \, \sigma_x  - \sqrt{{\kappa}/{s}} \, \xi_2\\
    \partial_t \vec \sigma &= - \vec \sigma \times \vec H(x)
    \end{array}
    \right.,
\end{align}
where $\vec H(x) := (2\lambda\, x,0,\omega_\rms) $ and $\xi_1$, $\xi_2$ are independent real Gaussian white noises with zero mean and variance $\langle\xi_i(t) \xi_j(t')\rangle = \delta_{ij} \delta(t-t')$~\cite{Buchhold_2018}. The unit length of the top is clearly conserved by the dynamics, $\sigma = 1$.
Let us stress that the sources of quantum fluctuations are twofold: the coupling to the dissipative environment and the quantum uncertainty in the initial condition.
We sample the initial conditions from the large-$s$ asymptotic expression of the Wigner function associated with $\rho_0$,
\begin{align} \label{eq:W0_xp}
     \frac{1}{2\pi} W_0(x,p, \vec \sigma) &=  \frac{s^2}{\pi^2} \rme^{- s(x^2 + p^2  + \sigma_\perp^2)}  \delta(\sigma-1),
\end{align}
where $\sigma_\perp^2 := \sigma_x^2 + \sigma_y^2$.
The numerical results are obtained in the resonant case $\omega_\rmc = \omega_\rms =: \omega_0$. The SDEs~(\ref{eq:eom_x_p_sigma}) are integrated by means of the stochastic Heun scheme, which ensures strong convergence of order $\rmd t$ to the solution~\cite{Bogoi_2023,Mannella_2025}. Details are provided in Sec.~\ref{app:Heun} of the SM.

In the thermodynamic limit, $s \to \infty$, one recovers the purely dissipative classical dynamics.
In the absence of noise, Eqs.~(\ref{eq:eom_x_p_sigma}) have two $\mathbb{Z}_2$-symmetric fixed points corresponding to the normal phase, located at $x^\ast=p^\ast=0$ with $\sigma_z^\ast=+1$ (north pole) and $\sigma_z^\ast=-1$ (south pole). The north pole is always unstable, while the south pole is stable for weak to moderate values of $\lambda$ until it loses stability via a pitchfork bifurcation at $\lambda=\lambda_\rmc$, branching out to two extra stable fixed points corresponding to the superradiant phase and located at 
\begin{align}
\left\{
\begin{aligned}
x^* &= \frac{\omega_\rms}{2\lambda}\sqrt{(\lambda/\lambda_\rmc)^4 - 1}\\
p^* &= \frac{\kappa}{\omega_\rmc}\,x^*
\end{aligned}
\quad
\begin{aligned}
\sigma_x^* &= -\sqrt{1-(\lambda_\rmc/\lambda)^4}\\
\sigma_y^* &= 0\\
\sigma_z^* &= -(\lambda_\rmc/\lambda)^2
\end{aligned}
\right.,
\end{align}
and its image under $\mathbb{Z}_2$: $x^* \mapsto -x^*$, $p^* \mapsto -p^*$ and $\sigma_{x,y}^* \mapsto  -\sigma_{x,y}^*$. In Fig.~\ref{fig:attractor}~(a), we illustrate a representative classical deterministic trajectory of the spin steadily converging to one of the superradiant fixed points.

At finite $s$, the presence of noise allows solutions to transition from one branch to another, thereby effectively restoring the $\mathbb{Z}_2$ symmetry. This is illustrated in Fig.~\ref{fig:attractor}~(b), where a single stochastic trajectory spends a sizable fraction of time in the vicinity of each of the superradiant fixed points, with intermittent `flights' passing through the vicinity of the south pole.
This generic phenomenon has been dubbed the destruction of a pitchfork bifurcation by additive noise~\cite{Crauel_1998}.
Here, it entails the rounding of the dissipative phase transition into a finite dimensional crossover.

\paragraph*{Random strange attractor.}
In the context of noisy dynamics, trajectories are intrinsically dependent on the noise realization.
Yet, attractors, which are inevitably promoted to random variables, can be meaningfully defined via the pullback dynamics (see, \textit{e.g.}, Refs.~\cite{Scheutzow_2002,Crauel_2015}).
This consists in starting the dynamics at time $t_0$ from a cloud of $N_0$ initial conditions sampled from the stationary distribution, and propagating them under the \emph{same} noise history to a final time $T$.
The pullback attractor is obtained as the limit of the final distribution as $t_0 \to -\infty$.
In Fig.~\ref{fig:attractor}~(c), we show a single realization of the pullback attractor for $\lambda>\lambda_\rmc$. For clarity, only the projection onto the spin sector is shown; the corresponding bosonic sector, together with additional details, are provided in Sec.~\ref{app:sec:pb-attractor} of the SM. The resulting set is compact yet extended in phase space, which we identify as a random strange attractor.
To substantiate this, we examine one of the hallmarks of strange attractors, namely their non-integer (fractal) dimension. Specifically, we compute the box-counting dimension of the attractor shown in Fig.~\ref{fig:attractor}~(c) by counting the number $N(\varepsilon)$ of boxes of linear size $\varepsilon$ required to cover the attractor in the four-dimensional embedding phase space. As shown in Fig.~\ref{fig:attractor}~(d), the scaling relation $\log N(\varepsilon) \sim D \log(1/\varepsilon)$ holds over a range of scales, yielding an estimated fractal dimension $D \approx 1.6$.

\begin{figure}
    \centering
    \includegraphics[width = \linewidth]{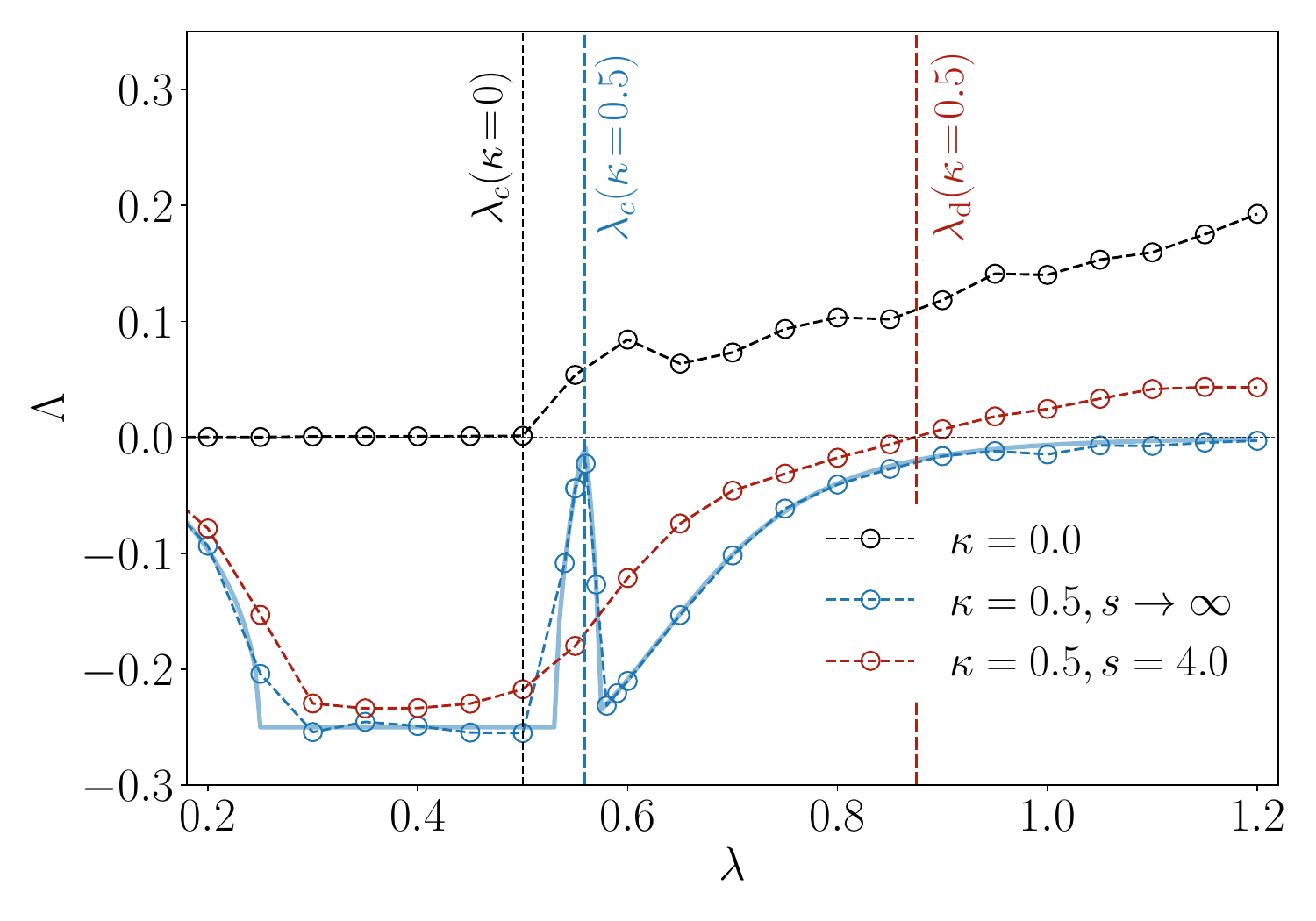}
    \caption{
    Effects of the dissipative environment on the Lyapunov exponent $\Lambda$. 
    Closed Dicke model, $\kappa = 0$ (black): the dynamics are chaotic $\Lambda > 0$ for spin-boson interactions $\lambda \gtrsim \lambda_\rmc $. 
    Classical dissipative Dicke model, $\kappa = 0.5$, $s\to \infty$ (blue): non-chaotic dynamics: $\Lambda \leq 0$ for all $\lambda$, matching the exact prediction from linearized dynamics (solid blue).
    Stochastic Dicke model, $\kappa = 0.5$, $s=4$ (red): finite-spin quantum fluctuations restores chaotic dynamics $\Lambda > 0$ for $\lambda  > \lambda_\rmd >\lambda_\rmc $.
    Parameters: $\omega_\rms = \omega_\rmc =1$ (10 initial states and 50 noise histories per initial state).
    }
    \label{fig:Lyapunov}
\end{figure}

\paragraph*{Lyapunov exponent.}
To quantify the chaoticity of this random strange attractor, we now characterize the Lyapunov exponent, which measures the asymptotic rate at which initially nearby trajectories separate in phase space,
\begin{equation}
     \Lambda := \lim_{T\to\infty} \frac1T \log  \left| \frac{\delta x(T)}{\delta x(0)}\right|.
\end{equation}
In principle, $\Lambda$ can be extracted from a single sufficiently long trajectory that explores the entire accessible phase space~\cite{Arnold_1998}.
In practice, however, the numerical integration of Eqs.~(\ref{eq:eom_x_p_sigma}) can only be carried out reliably up to a finite time $T$, making the above quantity a stochastic quantity. To alleviate this, we further average over the quantum fluctuations (initial conditions and noise realizations). Computational details are provided in Sec.~\ref{app:sec:lyapunov} of the SM.
The results are presented in Fig.~\ref{fig:Lyapunov}.
For the closed system ($\kappa=0$), energy is conserved and we sample the initial conditions from the energy shell $\mathcal{E} := \mathrm{Tr}[H \rho_0]/s + \varepsilon$ with the offset $\varepsilon = 10^{-2}\omega_\rmc$ corresponding to initial conditions slightly away from the south-pole fixed point.
We recover established results~\cite{Emary_PRL_2003,Emary_PRE_2003,Hirsch_2016}: the transition from regular to chaotic dynamics is marked by the onset of a positive Lyapunov exponent when the spin-boson coupling exceeds a dynamical threshold $\lambda_\rmd$ which coincides with the thermodynamic critical point, $\lambda_\rmd=\lambda_\rmc$.
The introduction of finite dissipation ($\kappa>0$) in the limit $s\to\infty$ stabilizes the classical fixed point~\cite{Santos_2026}. To accelerate convergence for $\lambda>\lambda_\rmc$, we initialize the dynamics in the vicinity of one of the stable superradiant fixed points. As expected, the Lyapunov exponent remains non-positive for all couplings. Notably, our numerical results are in excellent agreement with the exact predictions obtained from linearizing the dynamics around the stable fixed point (see solid blue in Fig.~\ref{fig:Lyapunov}), providing a non-trivial benchmark of the simulations. Further details are given in Sec.~\ref{app:sec:classical} of the SM.
At finite $s$, we find that the resulting quantum fluctuations are responsible for the restoration of a positive Lyapunov above a dynamical threshold $\lambda_\rmd > \lambda_\rmc$. 
This main result constitutes a clear and quantitative indicator of the chaotic nature of the random attractor identified above and illustrated in Fig.~\ref{fig:attractor}.

\begin{figure}[t]
    \centering
    \includegraphics[width = \linewidth]{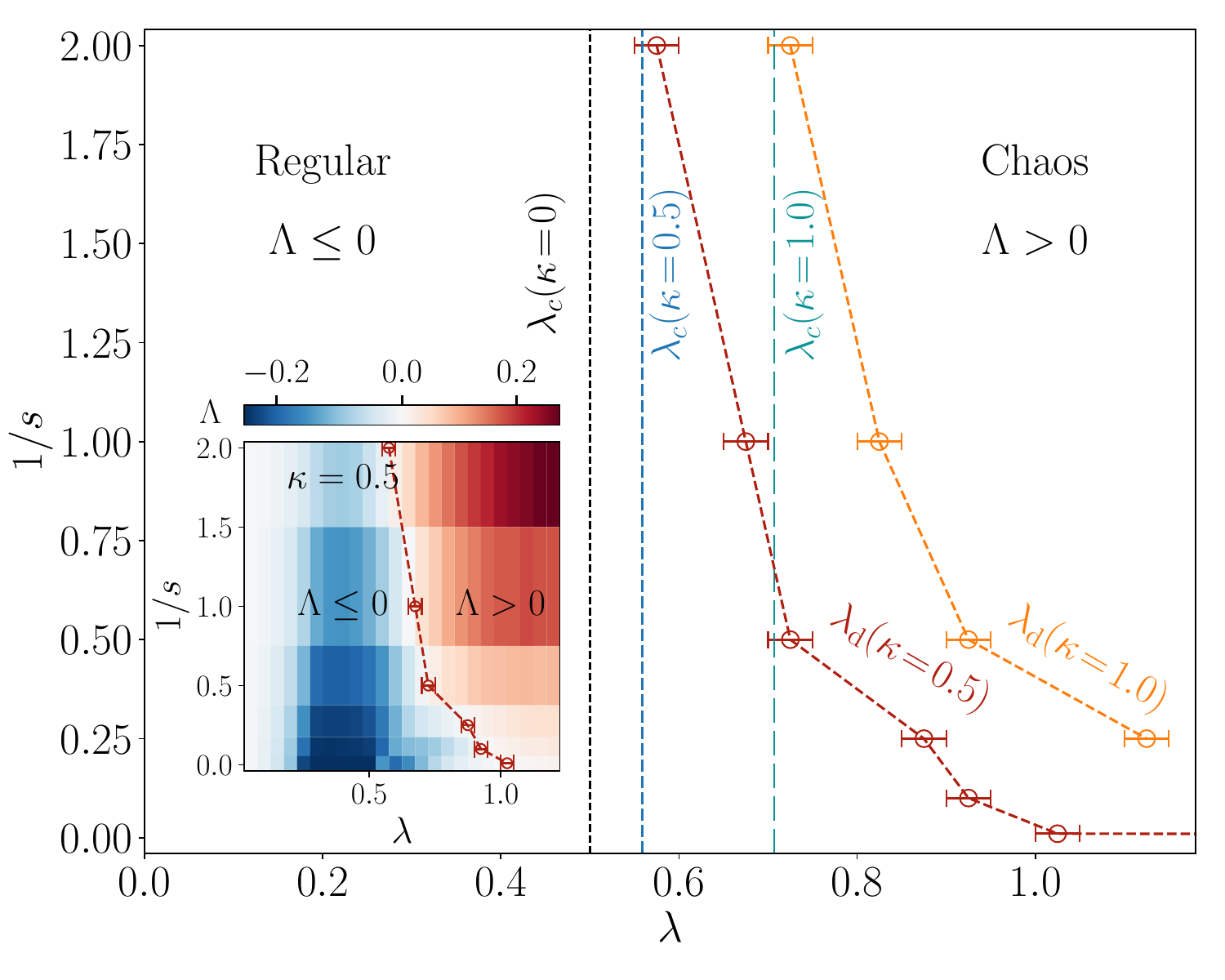}
    \caption{
    Dynamical phase diagram in the coupling $\lambda$ vs noise strength $1/s$ plane for several values of $\kappa$.
    The thermodynamical phase transitions at $s\to\infty$ are located at $\lambda_\rmc$ marked by the vertical lines.
    The dynamical thresholds $\lambda_\rmd$ separating non-chaotic ($\Lambda\leq 0$) and chaotic ($\Lambda>0$) regimes were extracted from data analogous to the finite-$s$ results shown in Fig.~\ref{fig:Lyapunov}.
    Error bars are set by the bin size in $\lambda$.
    Inset: heat map of the Lyapunov exponent for $\kappa=0.5$.
    Parameters: $\omega_\rms = \omega_\rmc =1$ (10 initial states and 50 noise histories per initial state).}
    \label{fig:phase-diagram}
\end{figure}

\paragraph*{Dynamical phase diagram.}
We now systematically investigate the role of quantum noise by computing the Lyapunov exponent $\Lambda$ as a function of $1/s$ and $\lambda$, determining the dynamical coupling $\lambda_\rmd$ above which $\Lambda$ becomes positive.
The resulting dynamical phase diagram is shown in Fig.~\ref{fig:phase-diagram}.
We find that sufficiently strong quantum fluctuations can always destabilize the classically regular dynamics in the superradiant phase, $\lambda\gtrsim\lambda_\rmc$, and generate a regime with a positive Lyapunov exponent.
We find that stronger coupling enhances susceptibility to noise-induced chaos.
In addition, increasing $\kappa$, which simultaneously controls dissipative damping and the strength of quantum bath fluctuations, is found to suppress chaotic behavior in the superradiant phase and to favor regular dynamics.
By contrast, in the normal regime, $\lambda \lesssim \lambda_\rmc$, the dynamics remain non-chaotic even in the presence of strong quantum noise.
In summary, the coupling $\lambda_\rmd$ marking the onset of chaos is a monotonously decreasing function of the noise strength, diverging in the classical limit $\lambda_\rmd (1/s \ll 1) \to \infty$, and bounded from below by the critical coupling $\lambda_\rmc$.

\paragraph*{Shear-induced chaos.}
We attribute this behavior to the distinct phase-space structures of the two phases. In the superradiant phase, noise-induced excursions can momentarily bring trajectories into the vicinity of the nearby south pole that is a classically unstable fixed point.
See, \textit{e.g.}, the stochastic trajectory in Fig.~\ref{fig:attractor}~(b). Since the linearized flow around this point has expanding directions, repeated visits contribute positively to the Lyapunov exponent. 
By contrast, the normal phase lacks a nearby unstable fixed point that could provide an analogous destabilizing mechanism and therefore remains regular. 
For the relatively weak noise strengths considered here, trajectories do not reach the vicinity of the unstable north pole.
This overall scenario could also be interpreted as the noise-induced amplification of the transient chaos observed at short times in the classical dissipative Dicke model~\cite{Santos_2026}, promoting it to steady-state chaos.
It is also interesting to note that, in the superradiant phase of the fully quantum closed Dicke model, the exponential growth of the out-of-time-order correlator (OTOC)--an annealed analogue of Lyapunov growth in which quantum fluctuations are averaged inside rather than outside the logarithm--is also governed by the instability of the south-pole saddle point~\cite{Cao_2020}.

This mechanism is closely related to the phenomenon of shear-induced chaos introduced by Lin and Young~\cite{Lin_2008}. It has been proven, with mathematical rigor, that the introduction of additive noise to a two-dimensional dynamical systems experiencing a Hopf bifurcation, \textit{i.e.} a single-parameter transition in which a focus loses stability giving rise to a limit cycle, can result in a random strange attractor with positive Lyapunov exponent on both sides of the bifurcation~\cite{Lin_2008,Ott_2008,Engel_2018,Engel_2019,Baxendale_2023,Engel_2023}. 
We refer the reader to Sec.~\ref{app:sec:shear-induced_chaos} of the SM for a brief pedagogical introduction to shear-induced chaos.
Although the superradiant transition is a pitchfork rather than a Hopf bifurcation, the dissipative Dicke model at $\lambda > \lambda_\rmc$ nevertheless exhibits similar spiraling motion once the dynamics approaches the vicinity of the superradiant fixed point. This originates from the fact that the linearized dynamics around the superradiant fixed points exhibit at least one pair of complex-conjugate eigenvalues, with an additional pair emerging at a node–focus transition occurring at $\lambda'' > \lambda_\rmc$. For $\kappa \ll \omega_0 := \omega_\rmc = \omega_\rms$, one finds $\lambda'' \simeq \lambda_\rmc + \kappa^2/(16\omega_0)$~\cite{Carmichael_2007,Oztop_2012}. Beyond this point, the angular velocity about the superradiant fixed point increases with $\lambda$, thereby providing the shear required by the Lin–Young mechanism.
Quantum noise then repeatedly drives trajectories away from the attracting fixed points and towards the unstable region around the south pole, where stretching occurs.
The resulting interplay between noise-induced reinjection, rotational motion, and local instability generates the random strange attractor observed in Fig.~\ref{fig:attractor}~(c).

\paragraph*{Conclusions and Outlook.} 
We have shown in an archetypal model of quantum optics that environmental quantum noise can bring chaos to an otherwise non-chaotic substrate. 
Here, we could attribute the chaos generation to the well-established mechanism of shear-induced chaos.
The important question is now to understand the relevance of this mechanism in quantum devices and quantum many-body problems in general.
A first step in that direction would be to study the case of the anisotropic or unbalanced dissipative Dicke model~\cite{Baden_2014,Zhang_2017,Aedo_2018,Tiwari_2023,Vivek_2025}, where the amplitude of the counter-rotating terms is different from that of the U(1)-symmetric terms. It hosts a rich dynamical phase diagram, with notably a line of Hopf bifurcations and chaotic regions that emerge from the simultaneous loss of stability of all the fixed points and a period-doubling route to chaos~\cite{Parkins_2020}. Investigating how these structures are modified by the addition of noise may uncover additional routes to chaos induced by quantum bath fluctuations.
Finally, while we have focused on the effects of an extrinsic bath, excited states of a large many-body system can, from a hydrodynamic viewpoint, effectively act as a bath for low-energy degrees of freedom. In the context of transmon qubits, such internal degrees of freedom have even been shown to induce chaos~\cite{Petrescu_2023}. It would therefore be interesting to clarify the similarities and differences between intrinsic and extrinsic baths in their respective roles in the emergence of chaotic dynamics.
\medskip

\paragraph*{Acknowledgments.}
We thank Xiangyu Cao and Vishal Vasan for insightful discussions.
M.K. and T.R. acknowledge support from the Department of Atomic Energy, Government of India, under Project No. RTI4001. 
We thank Kenan Henzelin for providing computational resources.
M.K. thanks the Institute of Physics at EPFL in Lausanne for their hospitality.
C.A. thanks the ICTS in Bangalore for their hospitality.

\let\oldaddcontentsline\addcontentsline
\renewcommand{\addcontentsline}[3]{}
\bibliography{references}
\let\addcontentsline\oldaddcontentsline%

\clearpage

\appendix
\onecolumngrid
\setcounter{figure}{0}
\setcounter{equation}{0}
\setcounter{page}{1}
\numberwithin{figure}{section}
\numberwithin{equation}{section}
\thispagestyle{empty}

\begin{center}
    {\bfseries\large
    \underline{Supplemental Material} \\[1ex]
    ``Chaos from quantum bath fluctuations''}\\[3ex]
    { Ilan Baud,$^1$ Tamoghna Ray,$^2$ Mahaveer Prasad,$^{3,4}$  Manas Kulkarni,$^2$ and Camille Aron$^{5,1}$}\\[1ex]
    {\small\itshape $^1$Institute of Physics, \'{E}cole Polytechnique F\'{e}d\'{e}rale de Lausanne (EPFL), 1015 Lausanne, Switzerland\\
    $^2$International Centre for Theoretical Sciences, Tata Institute of Fundamental Research, Bangalore 560089, India\\
    $^3$Science, Mathematics and Technology Cluster, Singapore University of Technology and Design, 487372 Singapore\\
    $^4$Centre for Quantum Technologies, National University of Singapore 117543, Singapore\\
    $^5$Laboratoire de Physique de l’\'{E}cole Normale Sup\'{e}rieure, ENS, Universit\'{e} PSL,\\ CNRS, Sorbonne Universit\'{e}, Universit\'{e} Paris Cit\'{e}, 75005 Paris, France} \\[1ex]
    {\small (Dated: \today)}
\end{center}

\tableofcontents

\setlength{\parindent}{0pt}
\setlength{\parskip}{5pt}

\section{Semiclassical approach}
\label{app:sec:TWA}
In this Section, we construct the Wigner phase-space representation of the dissipative Dicke model defined in Eqs.~(1) and~(2) of the main text, and we carefully take its large spin limit $s \gg 1$ to arrive at a semiclassical description of the dynamics which retains quantum fluctuations to the first non-trivial order in $1/s$.
It yields a set of coupled stochastic equations of motion, which we coin as the Stochastic Dicke Model.
Later, we take the classical limit $s\to\infty$, and review the known dynamical features of this model that are relevant to this work.

\subsection{Schwinger boson representation}
Let us first use the Schwinger boson representation of the spin degrees of freedom. It involves two auxiliary bosonic degrees of freedom, $b_\downarrow$ ($b_\downarrow^\dagger$) and $b_\uparrow$ ($b_\uparrow^\dagger$), and the mapping
\begin{equation}
S^z=\frac{n_\uparrow-n_\downarrow}{2}, \quad
S^+=b_\uparrow^\dagger b_\downarrow, \quad \text{with}\quad
n_\uparrow:=b_\uparrow^\dagger b_\uparrow, \qquad
n_\downarrow:=b_\downarrow^\dagger b_\downarrow,
\end{equation}
where the total number of Schwinger bosons $N_b := n_\uparrow + n_\downarrow$ is constrained: $N_b = 2s$.
The Hamiltonian reads
\begin{align}
    H = \omega_\rmc a^\dagger a +
    \frac12 \omega_\rms (b_\uparrow^\dagger b_\uparrow - b_\downarrow^\dagger b_\downarrow) +\frac{\lambda}{\sqrt{2s}} (a+a^\dagger) ( b_\uparrow^\dagger b_\downarrow +  b_\downarrow^\dagger b_\uparrow).
\end{align}
In this language, the initial conditions reads $\rho_0 = \op{0} \otimes \op{n_\uparrow=0,n_\downarrow=2s}$.
Importantly, the number of Schwinger bosons is conserved by the Hamiltonian dynamics as $[H,N_b] =0$, and the same applies to our Lindblad dynamics. We can therefore lift the Schwinger constraint: it is dynamically enforced provided it is satisfied by the initial state.

\subsection{Phase space formalism}
We briefly construct the phase space representation of quantum mechanics based on the Wigner representation. We refer the reader to Ref.~\cite{Polkovnikov_2010} for a pedagogical introduction.

\paragraph*{Wigner transform.}
Collecting the different bosonic degrees of freedom in a vector fashion,
\begin{equation}
\vec{\alpha} := (\alpha,\beta_{\uparrow},\beta_{\downarrow}) \quad \text{and} \quad \vec{\alpha}^{*} := (\alpha^{*},\beta_{\uparrow}^{*},\beta_{\downarrow}^{*}),
\end{equation}
the density matrix $\rho(t)$ is mapped to the Wigner function defined as
\begin{align}
    W(\vec\alpha,\vec\alpha^*;t) := \frac{1}{2^3} \int   \frac{\rmd \vec\xi^* \rmd\vec\xi}{(2 \pi \rmi)^3}\, \langle \vec\alpha- \vec\xi/2|\, \rho(t)\, | \vec\alpha + \vec\xi/2 \rangle \,
    \rme^{- (\vec\alpha^*-\vec\xi^*/2)\cdot(\vec\alpha+\vec\xi/2) - \vec\xi^*\cdot\vec\xi / 2},
\end{align}
where $|\alpha\rangle$ denotes the coherent state with $\alpha \in \mathbb{C}$.
We use the convention
\begin{align}
    |\alpha \rangle := \rme^{ a^\dagger \alpha} |0\rangle,
\end{align}
where $|0\rangle$ is the vacuum state with $\langle 0 | 0\rangle = 1$, $\langle \alpha | \alpha'\rangle = \exp(\alpha^*\alpha')$.
This convention corresponds to the completeness relation
\begin{align}
    \mathbb{I} = \int \frac{\rmd \alpha^* \rmd\alpha}{2\pi \rmi} 
    \rme^{-\alpha^*\alpha  }
    |\alpha\rangle \langle\alpha|,
\end{align}
where we note
\begin{align}
     \frac{\rmd \alpha^* \rmd\alpha}{2\pi \rmi} \equiv 
     \frac{\rmd \mathrm{Re}( \alpha)\, \rmd \mathrm{Im}( \alpha)}{\pi} = \frac{\rmd x \, \rmd p}{2\pi},
\end{align}
and where we introduced the quadratures $x := (\alpha+\alpha^*)/\sqrt{2}$ and $p :=\rmi (\alpha^*-\alpha)/\sqrt{2}$.

\paragraph*{Dynamics in the Wigner representation.}
In the Wigner representation of quantum mechanics, the Hamiltonian operator is mapped to a complex function of the phase space variables. Here, we have
\begin{align}
    H_W(\vec \alpha, \vec\alpha^*) = \omega_\rmc (\alpha^* \alpha - \frac{1}{2}) +
    \frac12 \omega_\rms (\beta_\uparrow^* \beta_\uparrow - \beta_\downarrow^* \beta_\downarrow) + \frac{\lambda}{\sqrt{2s}}(\alpha+\alpha^*) ( \beta_\uparrow^* \beta_\downarrow +  \beta_\downarrow^* \beta_\uparrow),
\end{align}
and the Lindblad master equation is mapped to the partial differential equation (PDE)
\begin{align}
  \rmi  \partial_t W & = \left[ \omega_\rmc \alpha^* \partial_{\alpha^*}  
  + \frac12\omega_\rms (\beta_\uparrow^* \partial_{\beta_\uparrow^*}  - \beta_\downarrow^* \partial_{\beta_\downarrow^*} ) 
  +  \frac{\lambda}{\sqrt{2s}} (\alpha+\alpha^\dagger) ( \beta_\uparrow^* \partial_{\beta_\downarrow^*}  + \beta_\downarrow^* \partial_{\beta_\uparrow^*} ) \right. \nonumber \\
  &
  \qquad \left.  +  \frac{\lambda}{\sqrt{2s}}( \beta_\uparrow^* \beta_\downarrow +   \beta_\downarrow^* \beta_\uparrow) \partial_{\alpha^*} - \mathrm{c.c.} \right] W + \rmi \kappa\Big[ 2 + \alpha \partial_\alpha +  \alpha^* \partial_{\alpha^*} + \partial_\alpha\partial_{\alpha^*} \Big] W \nonumber \\
  & \quad - \frac14 \frac{\lambda}{\sqrt{2s}} \left( \partial_{\alpha} -\partial_{\alpha^*}  \right) 
  \left(
  \partial_{\beta_{\uparrow^*}} \partial_{\beta_\downarrow} + 
  \partial_{\beta_{\downarrow^*}} \partial_{\beta_\uparrow} 
  \right) W. \label{app:eq:PDE_W}
\end{align}

\paragraph*{Initial conditions.}
The Wigner transform of a factorized state $\op{0} \otimes \op{n_\uparrow=0,n_\downarrow=2s}$ reads
\begin{align}
        8 (-1)^{2s} \rme^{-2(|\alpha|^2+|\beta_\uparrow|^2+|\beta_\downarrow|^2)} L_{2s}(4|\beta_\downarrow|^2)
\end{align}
where $L_n(x)$ is the Laguerre polynomial of order $n$.
The constraint on the Schwinger bosons, $N_b = 2s$, translates as $|\beta_\uparrow|^2 + |\beta_\downarrow|^2 = 2s$. Therefore, the initial condition $W(t=0)=W_0$ corresponding to the factorized state $\rho_0 = \op{0} \otimes \op{\downarrow}$ reads
\begin{align}
        W_0(\vec \alpha, \vec \alpha^*) = 8 (-1)^{2s} \rme^{-2|\alpha|^2-4s} L_{2s}(4|\beta_\downarrow|^2) \delta(|\beta_\uparrow|^2+|\beta_\downarrow|^2-2s) / Z_0,
\end{align}
with the normalization factor
\begin{align}
    Z_0 &:= \int \frac{\rmd \beta_\uparrow^*\rmd \beta_\uparrow\, \rmd \beta_\downarrow^*\rmd \beta_\downarrow}{(2\pi\rmi)^2} 4 (-1)^{2s} \rme^{-4s} L_{2s}(4|\beta_\downarrow|^2) \delta(|\beta_\uparrow|^2+|\beta_\downarrow|^2-2s) \\
    & = 
    4 (-1)^{2s} \rme^{-4s} 
     \int \frac{ \rmd \beta_\downarrow^*\rmd \beta_\downarrow}{2\pi\rmi} L_{2s
     }(4|\beta_\downarrow|^2) \Theta(2s-|\beta_\downarrow|^2)
     \\
    & =  (-1)^{2s} \rme^{-4s} \left[ L_{2s}(8s) -  L_{2s+1}(8s) \right].
\end{align}
It can be rewritten more compactly as
\begin{align} \label{eq:app_W0}
        W_0(\vec \alpha, \vec \alpha^*) = 8 \rme^{-2|\alpha|^2} \, L_{2s}(4|\beta_\downarrow|^2) \, \delta(|\beta_\uparrow|^2+|\beta_\downarrow|^2-2s) / Z_0',
\end{align}
with $Z_0' :=  L_{2s}(8s) -  L_{2s+1 }(8s)$.
In terms of the spin degrees of freedom $\vec \sigma = (\sigma_x, \sigma_y,\sigma_z)$, this translates to 
\begin{align} 
        W_0( \alpha,\alpha^*, \vec \sigma) &=  \frac{4}{\pi s} \rme^{-2|\alpha|^2} L_{2s}(4(s-\sigma_z)) \delta(\sigma-s) / Z_0',
\end{align}
where $\sigma^2 := \sigma_x^2 +\sigma_y^2 +\sigma_z^2$.
This expression is readily generalized to initial conditions corresponding to a product state of a bosonic coherent state $\alpha_0$ and a spin coherent state $\vec S_0$ polarized along an arbitrary direction $\vec u_0$. The initial density matrix is $\rho_0 = \op{\alpha_0} \otimes \op{\vec S_0}$
and the corresponding Wigner function reads
\begin{align} 
        W_0( \alpha,\alpha^*, \vec \sigma) &=  \frac{4}{\pi s} \rme^{-2|\alpha-\alpha_0|^2} L_{2s}(4(s-\sigma_\parallel)) \delta(\sigma-s) / Z_0',
\end{align}
where we decompose $\vec\sigma=:\sigma_\parallel \vec u_0+\vec\sigma_\perp$.

\subsection{Truncated Wigner Approximation}
We now consistently take the large-$s$ limit, and retain the first non-trivial $1/s$ corrections to the classical limit. This proceeds by first rescaling
\begin{align}
    \alpha \to \sqrt{s} \alpha\,, \quad
    \beta_\uparrow \to \sqrt{s}  \beta_\uparrow\,, \quad
    \beta_\downarrow \to \sqrt{s}  \beta_\downarrow\,,  \label{eq:app_rescaling}
\end{align}
and similar for their conjugate variables.

\paragraph*{Dynamics.}
At large $s$, after truncating the term in $1/s^2$ originating from the last line of Eq.~(\ref{app:eq:PDE_W}), we obtain a Fokker-Planck equation reading
\begin{align}
  \rmi  \partial_t W & = \left[ \omega_\rmc \alpha^* \partial_{\alpha^*}  
  + \frac12\omega_\rms (\beta_\uparrow^* \partial_{\beta_\uparrow^*}  - \beta_\downarrow^* \partial_{\beta_\downarrow^*} ) 
  +  \frac{\lambda}{\sqrt{2}} (\alpha+\alpha^\dagger) ( \beta_\uparrow^* \partial_{\beta_\downarrow^*}  + \beta_\downarrow^* \partial_{\beta_\uparrow^*} ) \right. \nonumber \\
  &
  \quad \left. \ +  \frac{\lambda}{\sqrt{2}} ( \beta_\uparrow^* \beta_\downarrow +   \beta_\downarrow^* \beta_\uparrow) \partial_{\alpha^*} - \mathrm{c.c.} \right] W + \rmi \kappa\Big[ 2 + \alpha \partial_\alpha +  \alpha^* \partial_{\alpha^*} + \frac1s \partial_\alpha\partial_{\alpha^*} \Big] W.
  \label{eq:app_LindbladTWA}
\end{align}
In this semiclassical approach, we retain two sources of quantum fluctuations: 
\begin{enumerate}
\item those encoded in the initial Wigner function of the PDE (see below), 
and 
\item the quantum bath fluctuations arising from the terms proportional to $\kappa/s$. 
\end{enumerate}
Consequently, in the absence of a quantum bath ($\kappa=0$), the resulting dynamics in Eq.~(\ref{eq:app_LindbladTWA}) are purely classical.
The subleading $1/s^2$ terms neglected in Eq.~(\ref{app:eq:PDE_W}) capture fluctuations generated intrinsically by the unitary quantum dynamics. They were shown to mitigate chaos in the closed Dicke model ($\kappa=0$) by washing out the fine phase-space structures associated with classical chaotic motion~\cite{Altland_Haake_PRL_2012,Altland_Haake_2012}.

\paragraph*{Initial conditions.}
We compute the asymptotic expression for $W_0(\alpha,\alpha^*,\vec\sigma)$ in Eq.~(\ref{eq:app_W0}) at large $s$. Let us first rescale as of Eq.~(\ref{eq:app_rescaling}), momentarily dropping prefactors,
\begin{align}
        W_0(\vec \alpha, \vec \alpha^*) &\sim \rme^{-2s|\alpha|^2} \, L_{2s}(8s - 4s |\beta_\uparrow|^2) \, \delta(|\beta_\uparrow|^2+|\beta_\downarrow|^2-2).
\end{align}
Using the fact that $L_{2s}(8s - 4s |\beta_\uparrow|^2)$ is maximal at $|\beta_\uparrow|^2 = 0$, using $L_{n}(4n-2n\epsilon) \propto \exp(-n\epsilon)$ at large $n$ and fixed $n\epsilon$, we obtain~\cite{Polkovnikov_2009}
\begin{align}
        W_0( \alpha,\alpha^*, \vec \sigma) &\sim \rme^{-2s|\alpha|^2} \rme^{-s\sigma_\perp^2} \delta(\sigma-1),
\end{align}
where we used $\sigma_\perp^2 = (2-|\beta_\uparrow|^2)|\beta_\uparrow|^2 \simeq 2 |\beta_\uparrow|^2$.
Recomputing the denominator, we get, at large $s$,
\begin{align}
     W_0( \alpha,\alpha^*, \vec \sigma) &\simeq    \frac{2s^2}{\pi} \rme^{-2s|\alpha|^2} \rme^{-s\sigma_\perp^2} \delta(\sigma-1),
\end{align}
or, in the $x-p$ representation of the boson:
\begin{align} \label{app:eq:W0_xp}
\frac{1}{2\pi}     W_0( x,p, \vec \sigma) &\simeq   \frac{s^2}{\pi^2} \rme^{-s (x^2 + p^2+ \sigma_\perp^2)} \delta(\sigma-1).
\end{align}
This expression is readily generalized to initial conditions corresponding to a product state of a bosonic coherent state $\alpha_0$ and a spin coherent state $\vec S_0$ polarized along an arbitrary direction $\vec u_{S_0}$: $\rho_0 = \op{\alpha_0} \otimes \op{\vec S_0}$. The corresponding Wigner function reads
\begin{align} 
\frac{1}{2\pi}     W_0( x,p, \vec \sigma) &\simeq   \frac{s^2}{\pi^2} \rme^{-s [(x-x_0)^2 + (p-p_0)^2+ \sigma_\perp^2]} \delta(\sigma-1),
\end{align}
where $x_0 := \sqrt{2} \, \mathrm{Re}(\alpha_0)$, $p_0 := \sqrt{2} \, \mathrm{Im}(\alpha_0) $ and we decompose $\vec\sigma=:\sigma_\parallel \vec u_{S_0}+\vec\sigma_\perp$.

\subsection{Stochastic Dicke Model}
In Eq.~(\ref{eq:app_LindbladTWA}), we recognize a Fokker-Planck equation of the form $\partial_t P = -\partial_x[f P] + D \partial_x^2 P$, where $P(x,t)$ is a time-dependent probability distribution, $f(x,t)$ is a force acting at the coordinate $x$, and $D>0$ is a diffusion constant.
This can be mapped to the overdamped Langevin equation $\partial_t x = f + \eta$ with the real Gaussian noise $\eta$ such that $\langle\eta(t) \rangle =0$ and $\langle \eta(t)\eta(t') \rangle = 2 D \delta(t-t')$.
The Fokker-Planck equation~(\ref{eq:app_LindbladTWA}) maps to a set of coupled stochastic differential equations that reads
\begin{align}
\everymath{\displaystyle}
\left\{
\begin{array}{rl}
    \rmi \partial_t \alpha &= \omega_\rmc \alpha + \frac{\lambda}{\sqrt{2}} (\beta_\uparrow^*\beta_\downarrow + \beta_\downarrow^*\beta_\uparrow) - \kappa \alpha +\sqrt{\frac{\kappa}{s}}\,\xi \\
    \rmi\partial_t \beta_\uparrow &= \frac12 \omega_\rms \beta_\uparrow + \frac{\lambda}{\sqrt{2}} (\alpha+\alpha^*) \beta_\downarrow \\
    \rmi\partial_t \beta_\downarrow &= - \frac12 \omega_\rms \beta_\downarrow + \frac{\lambda}{\sqrt{2}} (\alpha+\alpha^*) \beta_\uparrow 
\end{array}
    \right.,
\end{align}
and their complex-conjugate equations on the corresponding conjugated variables.
The complex Gaussian white noise is such that $\langle \xi(t) \rangle = 0$, $\langle \xi(t)\xi(t')\rangle = 0$, and $\langle\xi^*(t)\xi(t')\rangle = \delta(t-t')$.

\paragraph*{$\alpha-\vec\sigma$ coordinates.}
The complex Schwinger boson variables can now be replaced by real standard spin variables
\begin{align}
\sigma_x := \frac{\beta_\downarrow^*\beta_\uparrow + \beta_\uparrow^*\beta_\downarrow}{2}, \quad \sigma_y := \rmi \frac{\beta_\downarrow^*\beta_\uparrow - \beta_\uparrow^*\beta_\downarrow}{2}, \quad \sigma_z := \frac{\beta_\uparrow^*\beta_\uparrow - \beta_\downarrow^*\beta_\downarrow}{2}.
\end{align}
The equations of motion now read
\begin{align}\label{eq:eom_alphasigma}
\everymath{\displaystyle}
\left\{
\begin{array}{rl}
    \rmi \partial_t \alpha &= \omega_\rmc \alpha + 
    \sqrt{2} \lambda \sigma_x - \rmi\kappa \alpha +\sqrt{\tfrac{\kappa}{s}}\xi \\
    \partial_t \sigma_x &= - \omega_\rms \sigma_y \\
    \partial_t \sigma_y &=  \omega_\rms \sigma_x - \sqrt{2}\lambda (\alpha+\alpha^*) \sigma_z \\
    \partial_t \sigma_z & = \sqrt{2} \lambda  (\alpha+\alpha^*) \sigma_y
\end{array}
\right.,
\end{align}
with  $\langle\xi^*(t)\xi(t')\rangle = \delta(t-t')$.
Equivalent equations were derived via a different approach in Ref.~\cite{Buchhold_2018}.
Let us introduce the spin size $\sigma$ 
\begin{align}
    \sigma := \sqrt{\sigma_x^2 + \sigma_y^2 +\sigma_z^2} = \frac{1}{2} (|\beta_\uparrow|^2+|\beta_\downarrow|^2).
\end{align}
One may check that it is conserved by the dynamics: $\partial_t \sigma =0 $. Therefore, given the Schwinger constraint on the initial condition, we have the identity $ \sigma(t) = 1$ at all times.

\paragraph*{$x-p-\theta-\phi$ coordinates.}
To make better use of the fact that there are 4 degrees of freedom to solve, we now switch to a coordinate system where the boson is described by the real variables $x :=  (\alpha^* + \alpha)/\sqrt{2}$, $p:= \rmi (\alpha^*-\alpha)/\sqrt{2}$, and the spin is described by a classical top of unit length and spherical coordinates $\theta  \in [0,\pi]$ and $\phi \in [0,2\pi)$. In this coordinate system, the equations of motion read
\begin{align} \label{app:eq:eom_xpthetaphi}
\everymath{\displaystyle}
\left\{
\begin{array}{rl}
     \partial_t  x &=  \omega_\rmc \, p - \kappa \, x - \sqrt{\tfrac{\kappa}{s}} \, \xi_1\\[6pt]
    \partial_t p &= -\omega_\rmc \, x - 2 \lambda \, \sin\theta\cos\phi - \kappa \, p  - \sqrt{\tfrac{\kappa}{s}} \, \xi_2\\[6pt]
    \partial_t \theta &= -2 \lambda \, x \sin\phi\\[6pt]
    \partial_t \phi &= \omega_\rms - 2 \lambda  \, x \cot\theta \cos\phi
    \end{array}
\right.,
\end{align}
where the \textit{real} Gaussian white noise is such that $\langle\xi_i(t) \xi_j(t')\rangle = \delta_{ij} \delta(t-t')$ with $i,j = 1,2$.

\subsection{Pullback attractor}
\label{app:sec:pb-attractor}
In the study of non-autonomous dynamical systems, the definition of an attractor is a non-trivial issue that has led to the development of the notion of a \textit{pullback attractor}. This concept is particularly relevant for stochastic differential equations (SDEs), which are intrinsically non-autonomous: individual trajectories are conditioned on specific realizations of the noise $\xi$, rendering the standard deterministic notion of an attractor inadequate. The appropriate generalization is provided by the concept of pullback attractor~\cite{Crauel_2015}, which reduces to the usual deterministic attractor in the absence of noise and which we briefly review here for completeness.

A set $\mathcal{A}_\xi$ in phase space is said to be a pullback attractor if it is (i) compact, (ii) attracts any bounded set of initial conditions, and (iii) invariant under ``having started earlier". 
Crucially, in the presence of noise, $\mathcal{A}_\xi$ must be understood as random: each realization of the noise history $\xi(t)$ yields a distinct realization of $\mathcal{A}_\xi$. Point~(iii) is therefore to be interpreted as invariance with respect to propagating the starting point of the dynamics further back in time, keeping the rest of the noise history fixed.

In practice, we generate a single noise realization $\xi(t)$ from time  $t = t_0$ up to a final time $T$, where $t_0<0$ and $T > 0$ is chosen to be much larger than any characteristic timescale of the system. We then propagate a large ensemble of $N_0$ initial conditions, drawn from the stationary distribution, forward in time from $t_0$ to $T$ using the \emph{same} noise history for all trajectories. We record the set of phase-space coordinates at time $T$ and call it $\mathcal{A}_\xi(t_0)$. The pullback attractor $\mathcal{A}_\xi$ is then obtained as the limit $\mathcal{A}_\xi := \lim_{t_0 \to -\infty} \mathcal{A}_\xi(t_0)$, which we assess numerically by verifying that $\mathcal{A}_\xi(t_0)$ is insensitive to bringing $t_0$ further in the past.

\begin{figure}
    \centering
    \begin{minipage}{.49\linewidth}
        \includegraphics[width=0.49\linewidth]{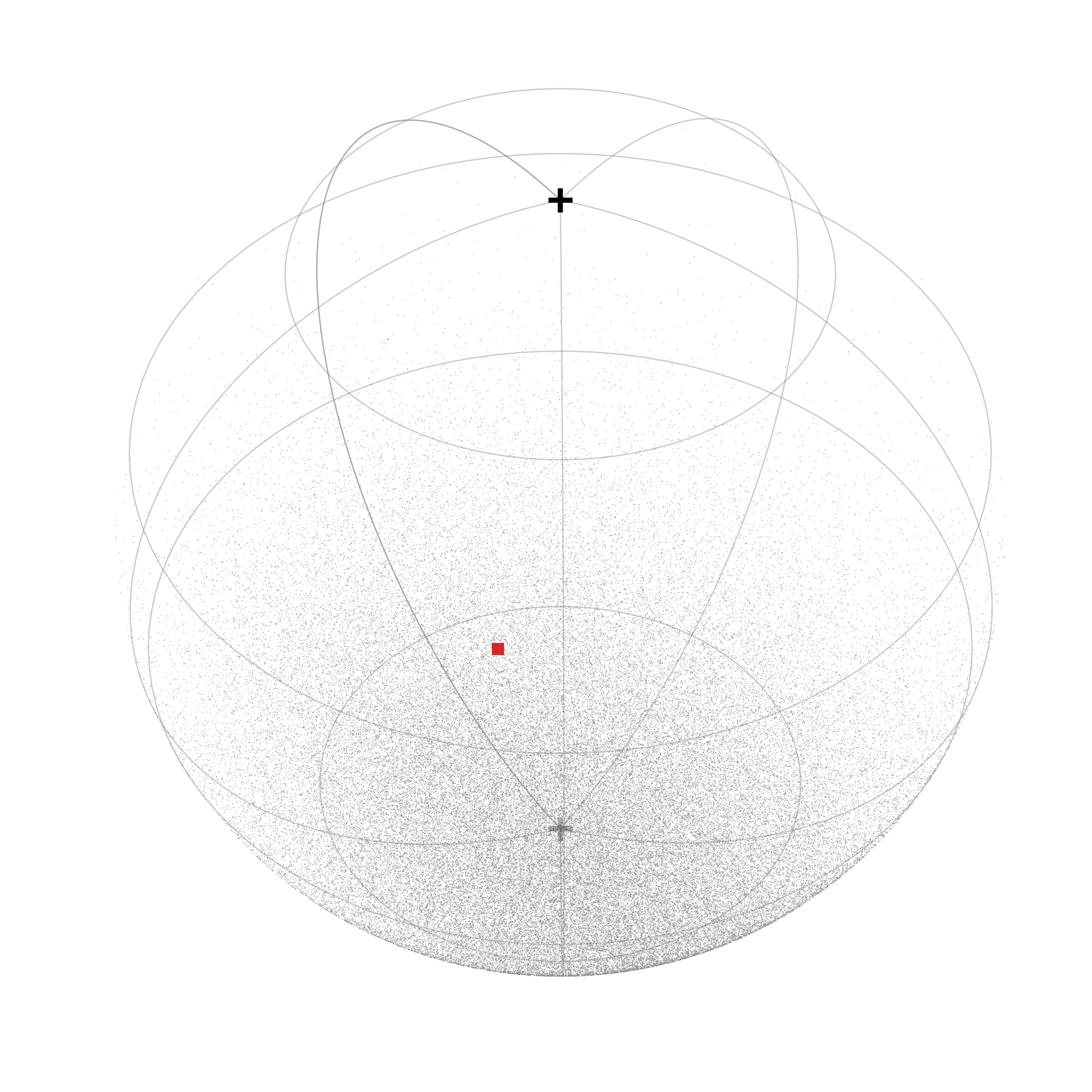}
        \includegraphics[width=0.49\linewidth]{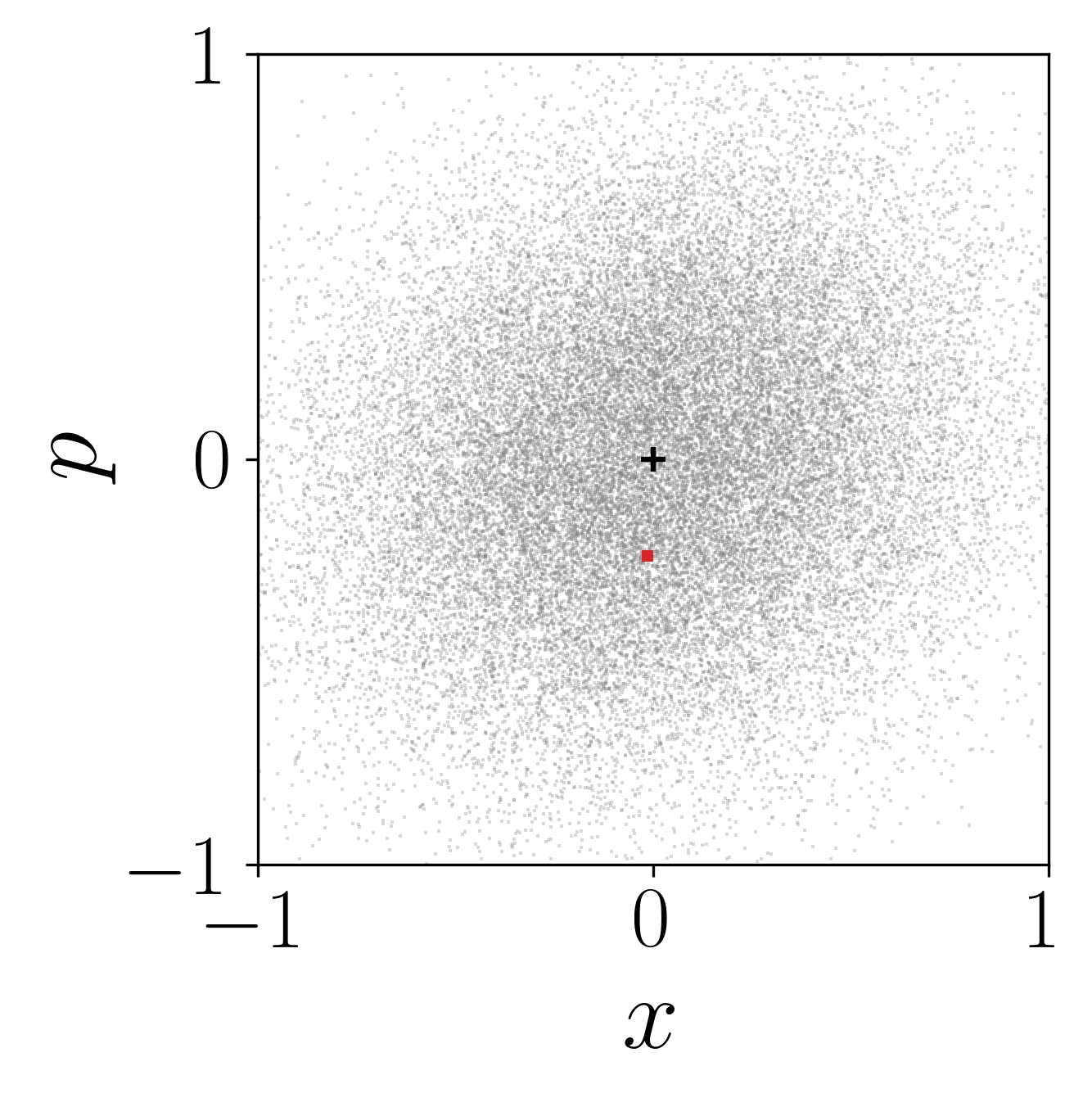}\newline
        {(a) $\lambda = 0.4 < \lambda_\rmc$}
    \end{minipage}
    \hfill
    \begin{minipage}{.49\linewidth}
        \includegraphics[width=0.47\linewidth]{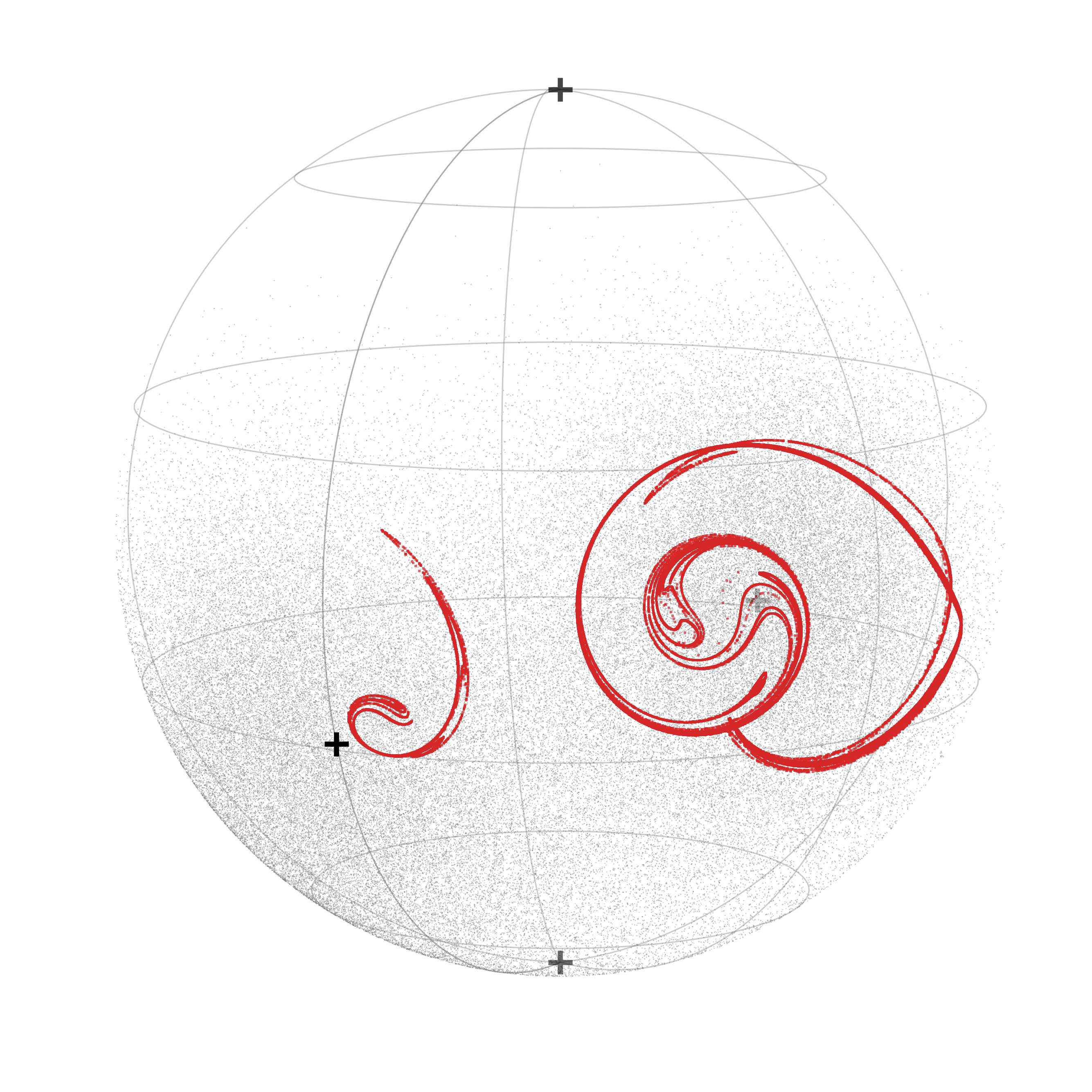}
        \includegraphics[width=0.49\linewidth]{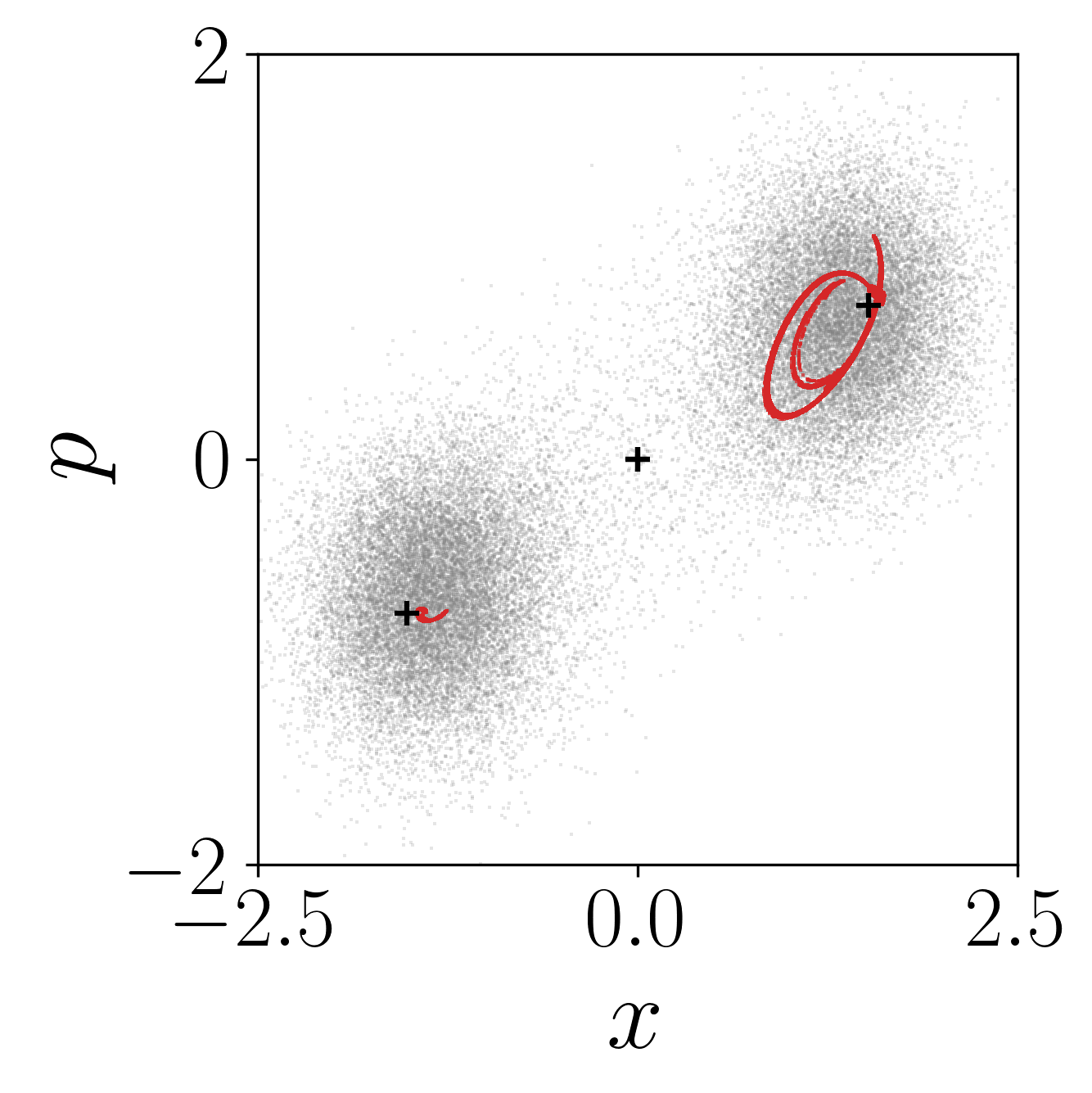}\newline
        {(b) $\lambda = 1 > \lambda_\rmc$}
    \end{minipage}
    \caption{
    Instances of the pullback attractor of the stochastic Dicke model (red), together with the ensemble of $N_0=10^5$ initial conditions (gray). Bloch-sphere representation of the spin and $(x,p)$ phase-space representation of the boson.
    (a) For $\lambda < \lambda_\rmc$, the pullback attractor reduces to a single point whose location depends on the noise realization.
    (b) At $\lambda = 1 > \lambda_\rmc$, the pullback attractor converges to an extended set in phase space, which we identify with a random strange attractor.
    Parameters: $\omega_\rmc = \omega_\rms = 1$, $\kappa = 0.5$, $s = 4$.
    }
     \label{app:fig:pb_dynamics}
\end{figure}

\begin{figure}
    \begin{minipage}{.32\linewidth}
        \includegraphics[width=\linewidth]{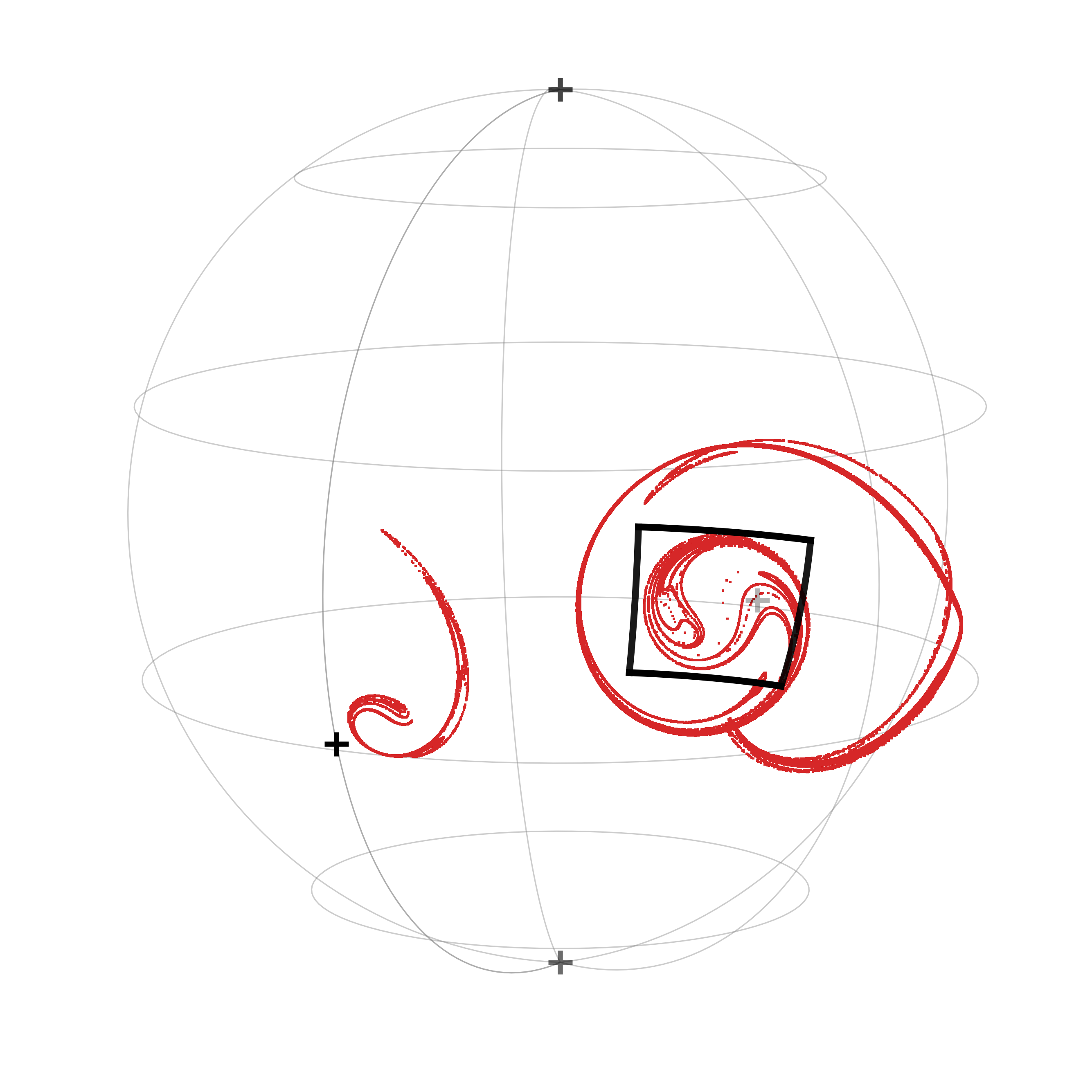}
    \end{minipage}
    \begin{minipage}{.32\linewidth}
        \includegraphics[width=\linewidth]{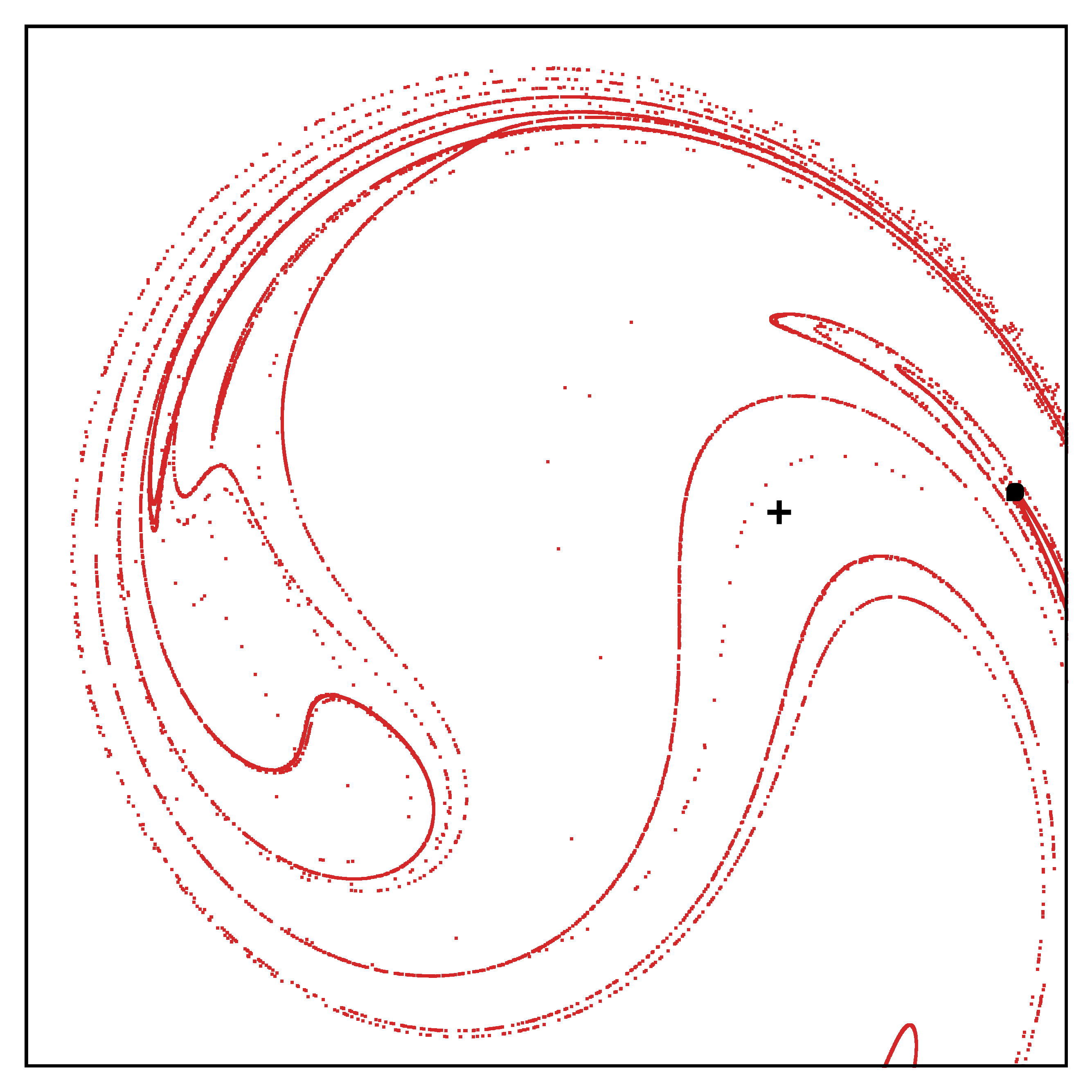}
    \end{minipage}
    \begin{minipage}{.32\linewidth}
        \includegraphics[width=\linewidth]{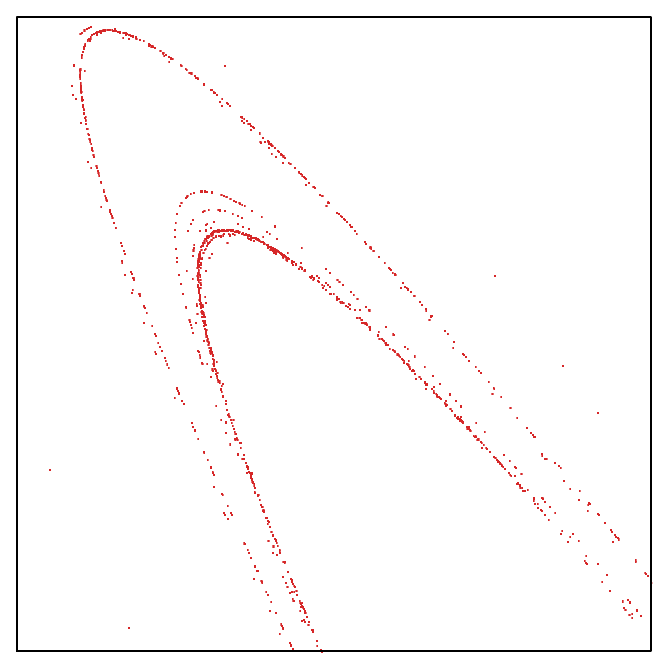}
    \end{minipage}
    \hspace{-182pt}\rule[10pt]{20pt}{1pt}\hspace{162pt}\textcolor{white}{.}
    
    \caption{Fractal-like structure of the pullback attractor in the superradiant regime. 
    (left) Random strange attractor as in Fig.~\ref{app:fig:pb_dynamics}.
    (middle) Zoom into a selected region of (left).
    (right) Further magnification of the region shown in (middle).
    Same parameters as in Fig.~\ref{app:fig:pb_dynamics}.
    }
    \label{app:fig:pb_fractal-like}
\end{figure}

In Fig.~\ref{app:fig:pb_dynamics}, we show realizations of the pullback dynamics of the stochastic Dicke model.
The sampling of the stationary measure with the $N_0$ initial conditions is plotted in gray. 
The pullback attractor $\mathcal{A}_\xi$ is plotted in red. 
In the normal regime, $\lambda < \lambda_\rmd$, the spatial extent of the pullback attractor collapses to zero, reducing to a single random point located near the south pole and $x=y=0$.
By contrast, in the superradiant regime, $\lambda > \lambda_\rmd$, $\mathcal{A}_\xi$ converges to an extended subset of phase space whose shape and support depend on the realization of the noise and are centered around one or both of the superradiant fixed points.
Notably, the overlap between $\mathcal{A}_\xi$ and the vicinity of the south pole, where the chaotic destabilization originates, appears to be negligible if not vanishing.
Furthermore, $A_\xi$ displays the hallmark features of a strange chaotic attractor: it forms a compact yet extended set in phase space, exhibiting winding fractal-like structures.
To further substantiate this claim, Fig.~\ref{app:fig:pb_fractal-like} shows the fractal-like patterns that emerge in the superradiant regime. In the main text, we additionally quantify this property by computing the fractal dimension of $\mathcal{A}_\xi$.

\subsection{Classical limit}
\label{app:sec:classical}
We review the $s\to \infty$ limit, which corresponds to dropping the noises in Eqs.~(\ref{app:eq:eom_xpthetaphi}), as well as starting from a deterministic classical initial condition.

\begin{figure}
    \centering
    \includegraphics[width=0.77\linewidth]{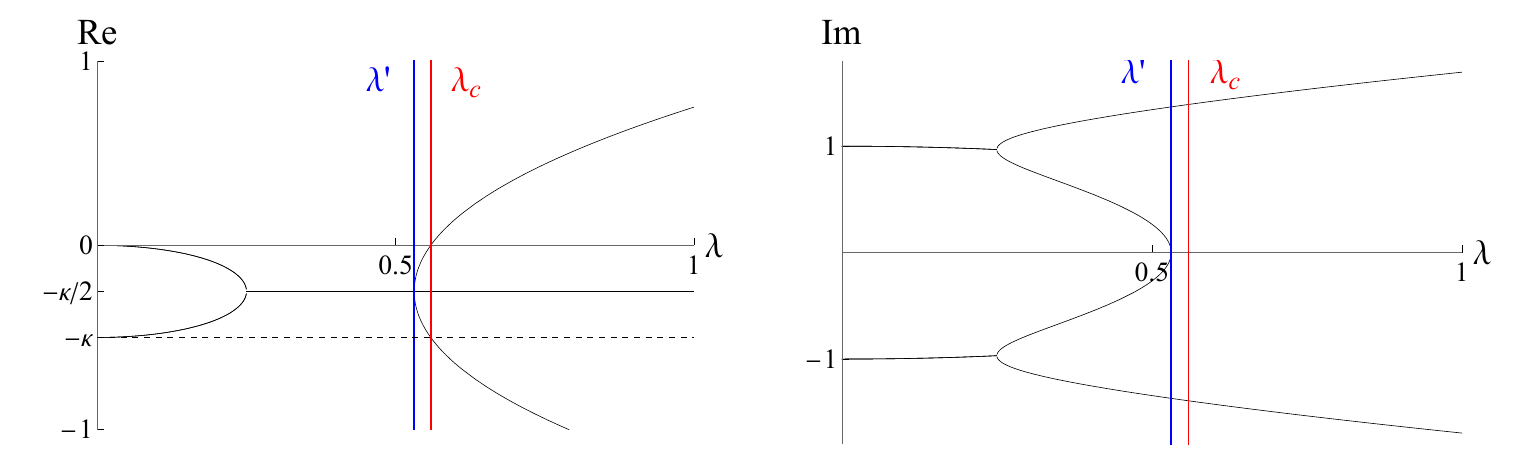}\newline
    (a) Normal south-pole fixed point.\vspace*{2mm}
    
    \includegraphics[width=0.77\linewidth]{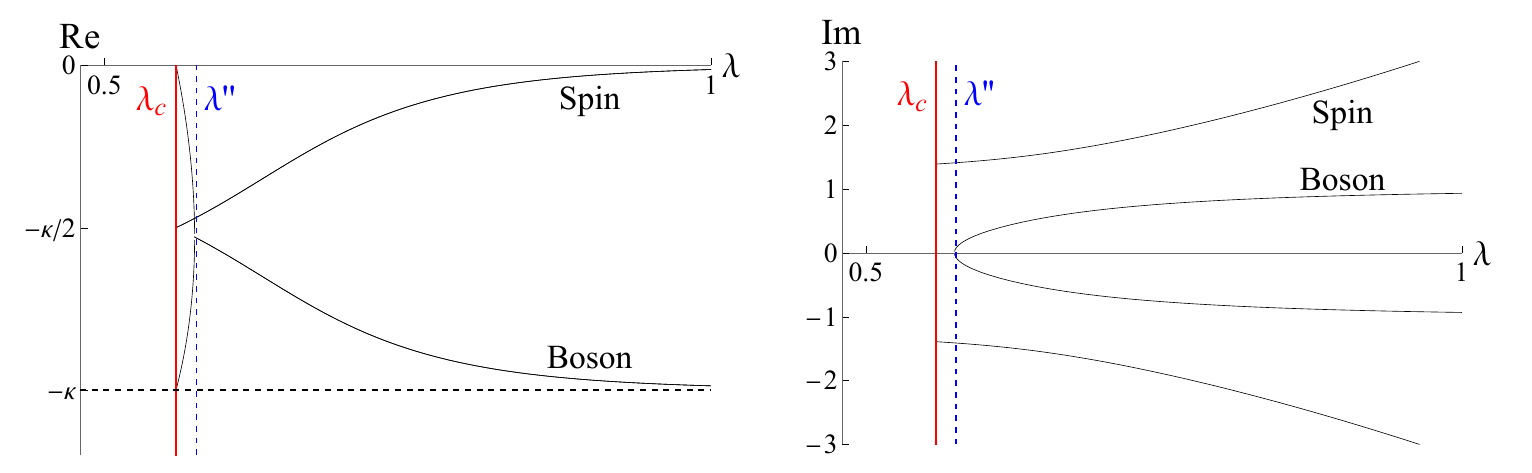}\newline
    (b) Superradiant fixed point.
    
    \caption{
    Linear stability: spectrum of the Jacobian in Eq.~(\ref{eq:Jacobian_mat}) at the classical fixed points.
(a) The south pole loses stability at $\lambda_\rmc$ due to a bifurcation triggered at $\lambda' < \lambda_\rmc$.
(b) The pair of stable broken-$\mathbb{Z}_2$ superradiant solutions appears at $\lambda_\rmc$ and undergoes a node-to-focus transition at $\lambda'' > \lambda_\rmc$, where two real negative eigenvalues merge into a complex-conjugate pair.
For $\lambda \gg \lambda_\rmc$, the spin and bosonic degrees of freedom decouple (as indicated by the branch labels), and the spin precession becomes virtually undamped. 
Parameters: $\omega_\rmc = \omega_\rms = 1, \kappa =0.5$.
    }
    \label{app:fig:Jac}
\end{figure}

\paragraph*{Fixed points.}
The classical fixed points of the deterministic system are given by solutions to the following set of equations 
\begin{align}
\everymath{\displaystyle}
\left\{
\begin{array}{rl}
    0 &= \omega_\rmc p^* - \kappa \,x^*\\
    0 &= -\omega_\rmc\, x^* - \kappa\,p^* - 2\lambda \sin\theta^*\cos\phi^*\\
    0 &= x^* \sin\phi^*\\
    0 &= \omega_\rms - 2\lambda \, x^* \cot\theta^*\cos\phi^*
\end{array}
\right..
\end{align}
Two trivial solutions are $x^*=p^*=0$, $\theta^*=0$ and $\theta^* = \pi$, where the oscillator is at rest and the top is pointing towards either the north pole or the south pole, where the azimuthal angle is irrelevant. They are both $\mathbb{Z}_2$-symmetric and correspond thermodynamically to the normal phase.

If $\theta^* \neq 0$ and $\theta^* \neq \pi$, the last equation ensures $x^* \neq 0$, $\theta^* \neq \pi/2$ and $\phi^* \neq \pm \pi/2$. 
Two non-trivial solutions appear when $\lambda > \lambda_\rmc$. They read
\begin{align}
\everymath{\displaystyle}
\left\{
\begin{array}{rl}
                x^* &= \tfrac{\omega_\rms}{2 \lambda} \sqrt{(\lambda/\lambda_\rmc)^4 - 1}\\
            p^* & = \tfrac{\kappa}{\omega_\rmc} x^* \\
           \cos\theta^* & = -(\lambda_\rmc/\lambda)^2\\
           \phi^* &= \pi
           \end{array}
\right.
\end{align}
and its image under the $\mathbb{Z}_2$-transformation $x^* \mapsto -x^*$, $p^* \mapsto -p^*$ and $\phi^* \mapsto  \phi^* - \pi$. These two non-trivial fixed points correspond to the superradiant phase where the weak-$\mathbb{Z}_2$ symmetry is spontaneously broken, yielding, notably, a non-vanishing expectation value of the electric field in the cavity, $\langle x^* \rangle \neq 0$.

\paragraph*{Linear stability analysis.}
The Jacobian associated with the linearization of the equations of motion is better expressed in the Cartesian basis $(x,p,\sigma_x, \sigma_y)$. We discard the $\sigma_z$ direction that is redundant because of the unit-length constraint. This reads 
\begin{align}
    J = 
    \begin{bmatrix}
-\kappa & \omega_\rmc & 0 & 0 \\
-\omega_\rmc & -\kappa & -2 \lambda & 0 \\
0 & 0 & 0 & -\omega_\rms \\
-2 \lambda \sigma_z & 0 & \omega_\rms & 0
    \end{bmatrix}.
    \label{eq:Jacobian_mat}
\end{align}
In Fig.~\ref{app:fig:Jac}, we plot the four eigenvalues of this Jacobian evaluated at the normal fixed point (south pole) and at the superradiant fixed points as a function of $\lambda$.
There is a transfer of stability from the former to the latter as $\lambda$ crosses $\lambda_\rmc$. 
This manifests itself as one of the Jacobian eigenvalues at the south pole develops a positive real part, whereas the eigenvalues at the superradiant solutions lie in the lower half plane as soon as the superradiant fixed point appears.

Interestingly, the destabilizing real part of the Jacobian spectrum originates from a prior bifurcation occurring at $\lambda' < \lambda_\mathrm{c}$. For $\kappa\ll\omega_0:=\omega_\rmc=\omega_\rms$, one finds $\lambda'\simeq \lambda_\rmc -\kappa^2/(8\omega_0)$~\cite{Carmichael_2007}.
For $\lambda < \lambda'$, the system exhibits a prolonged regime in which all eigenvalues have real part pinned to $-\kappa/2$. Since this is the only relevant dynamical scale entering the long-time evolution, the Lyapunov exponent correspondingly plateaus to $\Lambda = -\kappa/2$ throughout this extended interval.

Furthermore, there is a $\lambda''>\lambda_\rmc$ for which two real (negative) eigenvalues of the Jacobian at the superradiant fixed point merge into a pair of complex conjugate eigenvalues. This corresponds to a node-focus transition. 
For $\kappa\ll\omega_0$, one finds $\lambda''\simeq \lambda_\rmc +\kappa^2/(16\omega_0)$~\cite{Carmichael_2007}.
Both $\lambda'$ and $\lambda''$ merge with $\lambda_\rmc$ as $\kappa \to0$.
At large $\lambda \gg \lambda_\rmc$,  inspection of the Jacobian reveals the decoupling of the bosonic and spin degrees of freedom. In this limit and at the linear level, the boson spirals into its vacuum at a frequency $\omega_\rmc$ and a decay rate $\kappa$, while the spin rapidly rotates around its fixed point at a frequency $(\lambda/\lambda_\rmc)^2 \omega_\rms$, virtually undamped.

\section{Numerical methods}
\label{app:numerical_methods}
In this Section, we discuss the numerical methods used and the validity of the numerical data presented in this article.

\subsection{Stochastic Heun scheme}
\label{app:Heun}
Let us consider a set of coupled stochastic differential equations (SDEs) governing the dynamics of $N$ variables collected in the vector $\boldsymbol{x}:= (x_1,\ldots,x_N)$,
\begin{align}
\dot x_i = f_i(\boldsymbol{x}) + \sum_{j=1}^N\, g_{ij} \, \xi_j(t), \qquad i = 1, \dots, N,
\label{app:eq:generic_SDE}
\end{align}
where $f_i(\boldsymbol{x})$ denotes the deterministic drift, and $g_{ij}\geq 0$ are the coefficients of the diffusion matrix assumed to be independent of $\boldsymbol{x}$. The $\xi_j(t)$ are additive real Gaussian white noises with zero mean, $\langle \xi_j(t)\rangle = 0$, and correlations $\langle \xi_j(t)\xi_{j'}(t')\rangle = \delta_{jj'}\delta(t-t')$.

In the specific case of the stochastic Dicke model, $\boldsymbol{x}= (x,p,\theta,\phi)$, Eq.~(\ref{app:eq:generic_SDE}) reduces to Eq.~(\ref{app:eq:eom_xpthetaphi}), with $g_{ij}$ a diagonal matrix of entries $(\sqrt{\kappa/s}, \sqrt{\kappa/s}, 0, 0)$, and $\boldsymbol{\xi} = (\xi_1,\xi_2,0,0)$.

The stochastic Heun method is a second-order predictor–corrector scheme~\cite{Bogoi_2023,Mannella_2025}. To advance the state $\boldsymbol{x}(t)$ over a time step $\delta t$, the algorithm proceeds in two stages. First, a predictor step constructs an intermediate state $\tilde{\boldsymbol{x}}(t+\delta t)$ via an explicit Euler update,
\begin{align}
\tilde x_i(t + \delta t) = x_i(t) + \delta t \, f_i(\boldsymbol{x})+ g_i \sqrt{\delta t} \, Z_i,
\end{align}
where $g_i := g_{ii}$ and $Z_i$ are random variables sampled from a Gaussian distribution with zero mean and unit variance.

Second, a corrector step updates the variables by averaging drift and diffusion contributions evaluated at the initial and predicted states,
\begin{align}
x_i(t + \delta t) = x_i(t) + \frac{\delta t}{2}\Big[f_i(\boldsymbol{x}) + f_i(\tilde{\boldsymbol{x}})\Big] + g_i \sqrt{\delta t}\, Z_i.
\end{align}

The global error scales as $\mathcal{O}(\delta t)$, corresponding to strong order 1 convergence. In the noiseless limit ($g_i \to 0$ for all $i$), the scheme reduces to the standard second-order Runge–Kutta method. In our simulations, unless stated otherwise, we use a fixed time step $\delta t = 10^{-5}$ (in units of $1/\omega_c$).

\subsection{Lyapunov exponents}
\label{app:sec:lyapunov}
The sensitivity of the dynamics to infinitesimal perturbations of the initial conditions can be characterized by the $(4\times4)$ response matrix
\begin{equation}
    C_{ij}(t) := {\left| \frac{\delta x_i (t)}{\delta x_j(0)}\right|},
    \label{app:eq:C_mat}
\end{equation}
where $\boldsymbol{x} = (x, p, \theta, \phi)$ and the linearized perturbations $\delta x_i$ obey the following tangent-space dynamics
\begin{equation}
    \partial_t \delta  x_i(t) = J_{ij}(t) \, \delta x_j(t).
\end{equation} 
The Jacobian matrix $J(t) \equiv J({\boldsymbol  x}(t))$ reads 
\begin{align}
    J(t) = 
    \begin{bmatrix}
        - \kappa & \omega_\rmc & 0 & 0 \\
        -\omega_\rmc & -\kappa & -2\lambda\cos\theta\cos\phi & 2\lambda\sin\theta\sin\phi \\
        -2\lambda \sin\phi & 0 & 0 & -2\lambda x \cos\phi\\
        -2\lambda \cot\theta\cos\phi & 0 & 2\lambda x \csc^2\!\theta \cos\phi & 2\lambda x \cot\theta \sin\phi
    \end{bmatrix}.
\end{align}
The formal solution of the linearized dynamics is
\begin{align}
\delta {\boldsymbol  x} (t) = \mathtt{T} \rme^{\int_0^t \rmd s \, J(s)} \, \delta {\boldsymbol  x}(0),
\end{align}
where $\mathtt{T}$ denotes the time-ordering operator.
Discretizing time with step $\delta t = t/N_t$ where $N_t$ is a large integer, one obtains
\begin{align}
\delta {\boldsymbol  x} (t) = \mathtt{T}  \prod_{n=0}^{N_t-1} \Big[1+J(n \delta t)\delta t\Big]\, \delta {\boldsymbol  x}(0) + \mathcal{O}(\delta t^2),
\end{align}
where the product is the $4\times4$ matrix product.
Consistently with the stochastic Heun scheme described in Sec.~(\ref{app:Heun}), we neglect $\mathcal{O}(\delta t^2)$ terms. The response matrix can therefore be computed as
\begin{align}
    C_{ij}(t) 
    &= {\left| \left[ \mathtt{T}  \prod_{n=0}^{N_t-1} \Big[1+J(n\delta t)\delta t\Big]\right]_{ij} \right|}.
\end{align}

\begin{figure}
    \centering
    \includegraphics[width=0.7\linewidth]{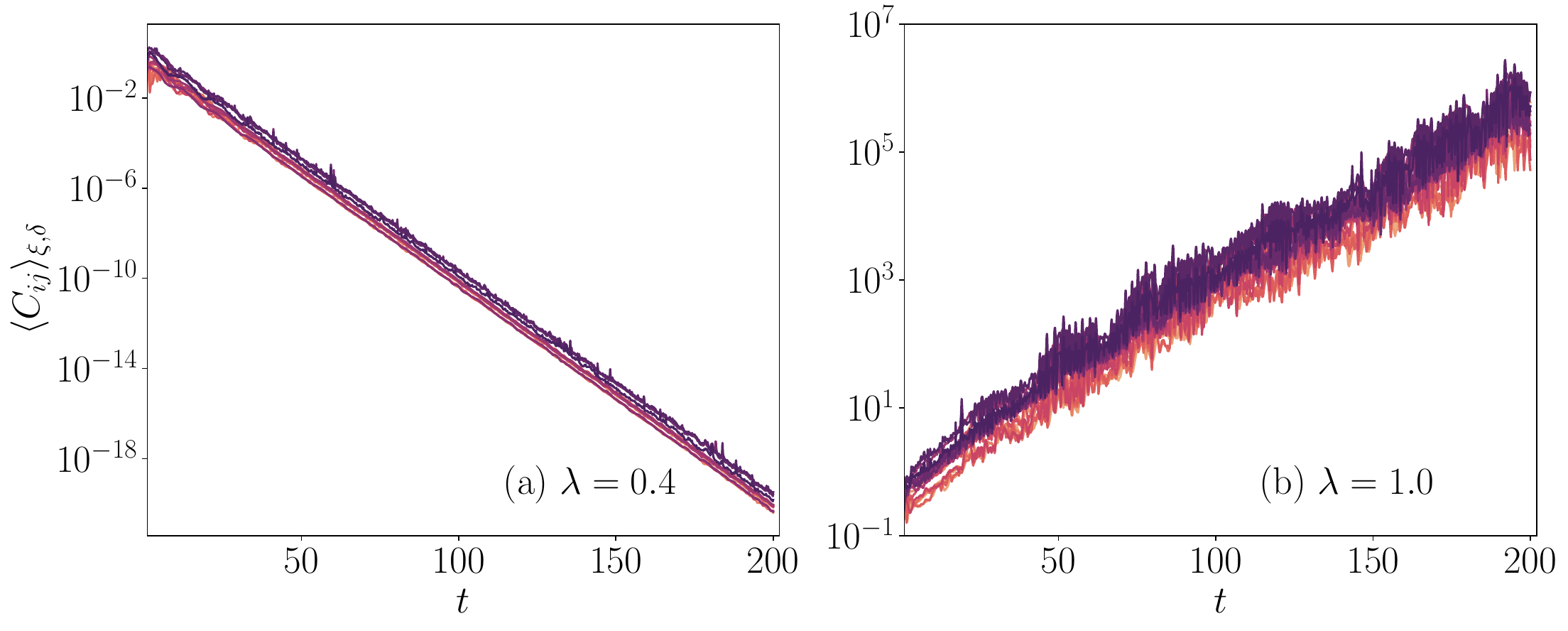}
    \caption{Different elements of the matrix $C$ in Eq.~(\ref{app:eq:C_mat}) are plotted as a function of time $t$ indicated by different colors. The Lyapunov exponent, extracted from the slope in the log-linear plot, is the same for all elements, both in the (a) regular regime ($\lambda = 0.4$) and in the (b) chaotic regime ($\lambda = 1.0$). Parameters: $\omega_\rmc = \omega_\rms = 1$, $\kappa = 0.5$, and $s = 4.0$.}
    \label{fig:C_ii_matrix}
\end{figure}

Lyapunov exponents quantify the exponential separation rate of infinitesimally close trajectories. In the present context, this naturally leads to a matrix of Lyapunov growth rates,
\begin{equation}
    \Lambda_{ij} := \lim_{t\to\infty} \frac1t \Big\langle   \log C_{ij}(t) \Big\rangle,
\end{equation}
where $\langle \ldots \rangle$ denotes the (quenched) average over both the quantum uncertainty in the initial state and the average over quantum bath fluctuations entering the dynamics.
We emphasize that the matrix $C_{ij}$ probes the response to infinitesimal perturbations applied along different but fixed phase-space directions. As such, it only provides access to the maximal Lyapunov exponent and not to the full Lyapunov spectrum. Under suitable ergodicity conditions, all matrix elements are expected to yield the same exponent $\Lambda_{ij} = \Lambda$ for all $i$ and $j$.
In Fig.~\ref{fig:C_ii_matrix}, we check this claim by showing the 16 components of $C_{ij}(t)$, which exhibit the same long-time exponential growth and therefore yield consistent estimates of the Lyapunov exponent.
In our study, we simply choose to extract the Lyapunov exponent from the $x$-$x$ component, $\Lambda \equiv \Lambda_{xx}$.

\begin{figure}
    \centering
    \includegraphics[width=\linewidth]{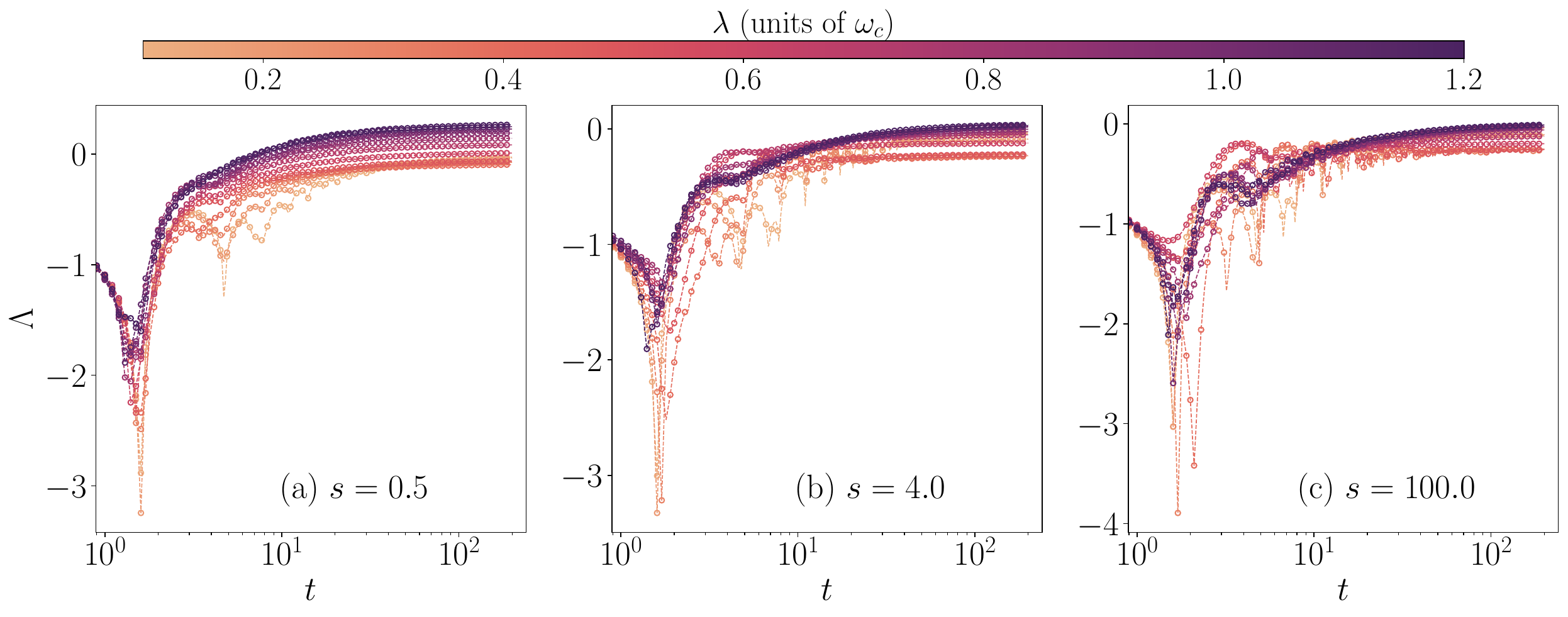}
    \caption{
    Numerically extracted Lyapunov exponent $\Lambda$ as a function of time $t$ for (a) $s = 0.5$, (b) $s = 4.0$, and (c) $s = 100.0$, for several values of the spin-boson coupling  $\lambda$ as indicated by the color bar. 
    Parameters: $\omega_\rmc = \omega_\rms = 1$, and $\kappa = 0.5$.}
    \label{fig:Lyap_v_t}
\end{figure}

In practice, we extract the Lyapunov exponent via a linear fit of $\langle \log C_{xx}(t)\rangle$ over the time window $[t/2,t]$, with $t$ chosen sufficiently large to reach the asymptotic regime.
To illustrate this procedure, we plot in Fig.~\ref{fig:Lyap_v_t} the extracted Lyapunov exponent as a function of time $t$, for different values of $\lambda$ and $s$. We consider dynamics up to $t = 200$, which is sufficient for the Lyapunov to reach a steady value.

\begin{figure}
    \centering
    \includegraphics[width=\linewidth]{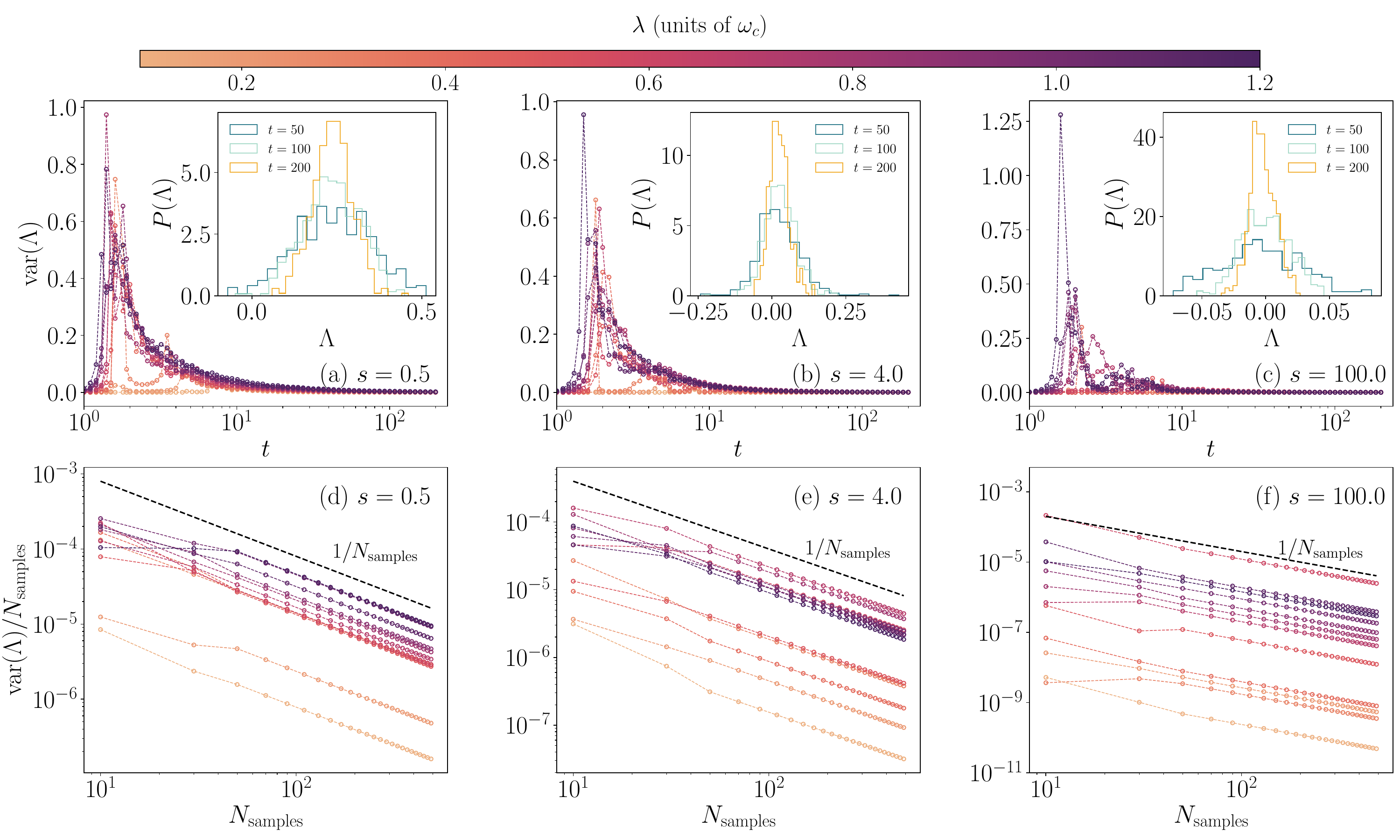}
    \caption{
    Top panel: variance of the distribution of Lyapunov exponents $p(\Lambda)$ arising from $N_{\rm samples} = 500$ realizations of initial conditions and noise, as a function of time $t$ for (a) $s=0.5$, (b) $s=4.0$, and (c) $s=100.0$, and for several values of $\lambda$ as indicated by the color bar.
    Insets of (a-c): distribution of Lyapunov exponents $p(\Lambda)$ extracted at selected times given in the key and for $\lambda=1.0$. 
    Bottom panel: variance of $p(\Lambda)$ as a function of the number of samples, at a fixed time $t = 200$ (in units of $1/\omega_\rmc$) for (d) $s = 0.5$, (e) $s = 4.0$, and (f) $s = 100.0$, and for several values of $\lambda$. Parameters: $\omega_\rmc = \omega_\rms = 1$, and $\kappa = 0.5$.
    }
    \label{fig:var_Lyap_v_t}
\end{figure}

Since different realizations of the quantum fluctuations (in the initial state and the quantum noise) yield different finite-time Lyapunov exponents, we also illustrate the corresponding distribution $p(\Lambda)$.
In Fig.~\ref{fig:var_Lyap_v_t}~(a-c), we show the time dependence of the variance of $\Lambda$ for several spin sizes. As time increases, the distributions become progressively narrower. The insets of Fig.~\ref{fig:var_Lyap_v_t}~(a-c) display the full distributions $p(\Lambda)$ at selected times.
In Fig.~\ref{fig:var_Lyap_v_t}~(d-f), we examine the convergence with the total number of samples, $N_{\rm samples}$, which includes averaging over both initial-state realizations and quantum-noise realizations $\xi$, for several spin sizes. 
We find that $p(\Lambda)$ converges towards a finite-width distribution as a function of $N_{\rm samples}$. Therefore, the estimation of $ \langle \Lambda \rangle$ converges as $1/\sqrt{N_{\rm samples}}$.
Together, these results demonstrate that $ \langle \Lambda \rangle$ can be reliably estimated.

\section{Shear-induced chaos}
\label{app:sec:shear-induced_chaos}
This Section is intended to be self-contained and can be read independently of the rest of the paper. We numerically reproduce selected results from a series of works in the mathematical literature~\cite{Lin_2008,Ott_2008,Engel_2018,Engel_2019,Baxendale_2023,Engel_2023}, which rigorously establish that a stable (non-chaotic) focus can be destabilized into a chaotic attractor by a time-dependent perturbation, leading in particular to a positive top Lyapunov exponent.
This setting further provides a useful benchmark for validating our numerical procedures, including Heun's integration scheme, pullback dynamics, box-counting dimension estimates, and Lyapunov exponent computations.
Combined with our perspective where quantum corrections to the classical dissipative Dicke model are precisely generating noise on top of stable-focus dynamics, this provides a concrete proof of concept that quantum-generated fluctuations can induce chaotic behavior.

\subsection{2D stochastic toy model}
Following Refs.~\cite{Engel_2018,Baxendale_2023,Engel_2023}, we consider the following 2D system of SDEs which realizes the normal form of a Hopf bifurcation subject to additive white noise,
\begin{align} \label{app:eq:hopf_eq}
\everymath{\displaystyle}
\left\{
\begin{array}{rl}
     \partial_t x &= \mu x - \omega y - (a x - b y) (x^2+y^2) + \sigma \, \xi_1 \\
    \partial_t y  &= \mu y + \omega x - (b x + a y)(x^2+y^2)  + \sigma \, \xi_2
\end{array}
    \right.,
\end{align}
with $\mu,\omega,b,\sigma \in \mathbb{R}$, $a>0$, and where $\xi_1(t)$ and $\xi_2(t)$ are real Gaussian white noises with zero mean and variance $\langle \xi_i(t) \xi_j(t') \rangle = \delta_{ij} \delta(t-t')$.

As it will become clearer below, the analogy between this 2D system and the stochastic Dicke model is as follows: the parameters $\mu$ and $b$ can be roughly seen as analogs of $-\kappa$ and $\lambda$, respectively. The noise strength $\sigma^2 \geq 0$ is the analog of $1/s$.

Without noise ($\sigma=0$): $\mu$ drives a supercritical Hopf bifurcation between a globally stable fixed point located at the origin ($\mu < 0$) to a stable limit cycle centered at the origin and with radius $\sqrt{\mu/a}$ ($\mu > 0$).
The parameter $a$ controls the local stability of the limit cycle, $\omega$ sets the angular velocity, while $b$ governs the shearing of the dynamics in the vicinity of the limit cycle.
Here, since the classical dissipative Dicke model does not have a limit cycle, but only a node focus at $\lambda > \lambda''$, we shall only consider the case $\mu < 0$, before the onset of the Hopf bifurcation. Similar results can be obtained at $\mu > 0$ without added difficulty.

For $\mu < 0$, the analysis of the eigenvalues of the Jacobian obtained by linearizing the dynamics reveals the existence of an extended region in phase space containing an expanding direction where one of the eigenvalues has a positive real part. Straightforward algebra yields an annulus centered at $x=y=0$ and bounded by the two radii
\begin{align} \label{app:eq:annulus}
  r_\mp :=  \frac{1}{\sqrt{3}C} \sqrt{2 K \mp \sqrt{K^2-3\Delta^2}},
\end{align}
where $C^2 := a^2 + b^2$, $K := a\mu+b\omega$, and $\Delta := a\omega-b\mu$. Although this local instability has little impact on the deterministic dynamics, for which all trajectories ultimately relax to the attractive fixed point at $x=y=0$, it plays a crucial role in the presence of noise, analogous to that of the south pole in the stochastic Dicke model.
Let us stress that, in the absence of noise, the late-time dynamics are entirely governed by this stable fixed point and are therefore globally non-chaotic.

In the presence of noise ($\sigma >0$), the invariant measure of the Fokker-Planck evolution associated with Eqs.~(\ref{app:eq:hopf_eq}) corresponds to a rotationally invariant Gibbs-like state of the form
\begin{align} \label{eq:stat_distrib}
    P_\infty(x,y) \propto \exp\left[ - \frac{V(x,y)}{\sigma^2} \right],
\end{align}
where the effective potential is
\begin{align}
V(x,y) = \frac{a}{2}   \left[ (x^2+y^2)-\mu/a\right]^2,
\end{align}
and $\sigma^2$ can be seen as an effective temperature.
While $\mu$ clearly drives a qualitative change in this invariant measure (from a 2D parabola to a Mexican hat potential), the invariant measure is unique and does not break rotational symmetry as long as $\sigma^2 > 0 $.

Remarkably, these statics bear no signature of the annulus introduced above in Eq.~(\ref{app:eq:annulus}). 
More generally, the deterministic drifts in Eqs.~(\ref{app:eq:hopf_eq}) do not derive from the potential $V(x,y)$ unless $b=\omega=0$: the dynamics cannot be written as relaxational dynamics \`a la Model A of the form $\partial_t x = -\partial_x V(x,y) + \sigma\, \xi_1(t)$ and $\partial_t y = -\partial_y V(x,y) + \sigma\, \xi_2(t)$. This means that while the statics can be put in the form of an equilibrium Gibbs state, the dynamics do not obey detailed balance.
More importantly, although the stationary distribution $P_\infty$ is independent of $b$ and $\omega$, these two non-conservative parameters are key to the shear-induced chaos mechanism.
This is supported by the inspection of Eq.~(\ref{app:eq:annulus}), as the existence of the annulus relies on a sufficiently large $K > 0$,  which requires $b \omega$ larger than $a |\mu|$. Consistent with this picture, it has been rigorously demonstrated that the noise-induced Lyapunov exponent $\Lambda$ grows as $b^{2/3}$ at large $b$ (keeping all other parameters fixed)~\cite{Baxendale_2023}.

\subsection{Random strange attractor}
\begin{figure}
    \centering
    \includegraphics[width=0.49\linewidth]{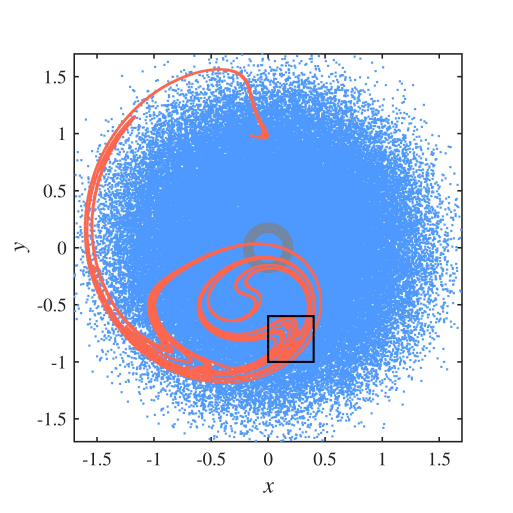}
    \includegraphics[width=0.46\linewidth]{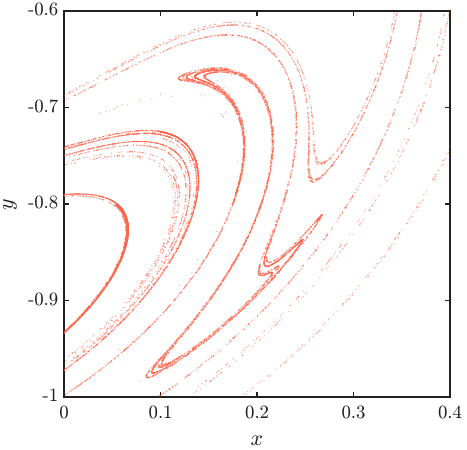} 
    \caption{Random strange attractor $\mathcal{A}_\xi$ induced by noise in the Eqs.~(\ref{app:eq:hopf_eq}).
    (left) The $N_0 = 10^5$ initial conditions (blue dots) are drawn from the stationary distribution in Eq.~(\ref{eq:stat_distrib}). 
    The gray annulus delimited by the radii $r_\mp$ in Eq.~(\ref{app:eq:annulus}) corresponds to the presence of an expanding direction.
    Pullback dynamics: The random attractor (red dots) is the set of final $\left(x(T),y(T)\right)$ where all trajectories experienced the \emph{same} noise history.
    (right) Zoom-in on a fractal feature of the strange attractor found on the left.
    Parameters: $\mu = -0.1$, $\omega  = a =  \sigma = 1$, $b = 20$ ($\delta t = 10^{-4}$, $T = 10$). 
    }
    \label{fig:appPullback}
\end{figure}
In the context of noisy dynamics, single trajectories depend on the realization of the noise.
The generalization of the concept of chaotic attractor to this noisy context relies on the so-called pullback dynamics.
This consists in starting the dynamics at time $t_0$ from a cloud of $N_0$ initial conditions that samples, say, the stationary distribution, and evolving them with the \emph{same} noise history $\xi_i(t)$ until a final time $T$.
The pullback attractor $\mathcal{A}_\xi$ is obtained as the limit of the final distribution when taking $t_0 \to -\infty$.
In Fig.~\ref{fig:appPullback}, we show the result of one realization of the pullback dynamics corresponding to Eqs.~(\ref{app:eq:hopf_eq}) in the node-focus regime, \textit{i.e.} $\mu < 0$.
We stress that the same noise history is used for all the trajectories.
The pullback attractor $\mathcal{A}_\xi$ is plotted in red. 
It is random in that it depends on the noise history. Yet, it is converged with respect to the initial time $t_0 \to - \infty$. 
Notably, the random attractor has the generic features of a chaotic strange attractor: a compact set that extends in phase space, with winding structures reminiscent of fractal dimensions.
Averaging over all possible noise histories brings back the stationary measure given in Eq.~(\ref{eq:stat_distrib}): $\langle A_\xi\rangle  = P_{\infty}$.

\subsection{Fractal dimension}
\begin{figure}
    \centering
    \includegraphics[width=0.35\linewidth]{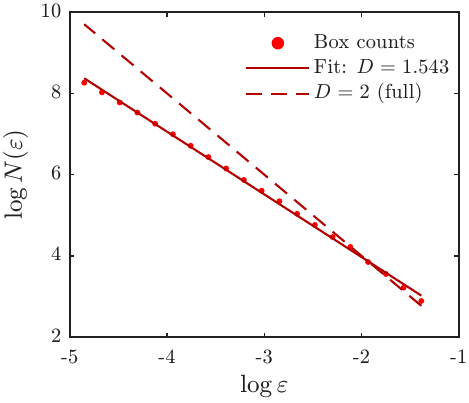}
    \caption{
    Fractal dimension of the random strange attractor in Fig.~\ref{fig:appPullback} estimated by box counting.
    Parameters: $\mu = -0.1$, $\omega  = a =  \sigma = 1$, $b = 20$ ($\delta t = 10^{-4}$, $T=10$). 
    }
    \label{fig:appBoxCounting}
\end{figure}
The fractal dimension of a strange attractor quantifies the degree to which the long-time dynamics fills the available phase space. Trajectories contract onto an invariant set of measure zero yet of non-trivial geometry, and the box-counting dimension characterizes the scaling of this set across length scales. Covering this invariant set with a grid of hypercubes of side length $\varepsilon$, one counts $N(\varepsilon)$, the number of boxes visited by the attractor.
For a self-similar set, this obeys the power law $N(\varepsilon) \sim \varepsilon^{-D}$ as $\varepsilon \to 0$, defining the box-counting dimension
\begin{equation}
    D := -\lim_{\varepsilon \to 0} \frac{\log N(\varepsilon)}{\log \varepsilon}.
\end{equation}
A non-integer value of $D$ is the hallmark of a strange attractor: the dynamics are confined to a set that is neither a smooth manifold nor a dense filling of phase space, reflecting the interplay between the stretching and folding of chaotic trajectories and the overall phase-space contraction.

In practice, we estimate the box-counting dimension $D$ of the random strange attractor $\mathcal{A}_\xi$ shown in Fig.~\ref{fig:appPullback} from the slope of $\log N(\varepsilon)$ versus $\log(1/\varepsilon)$ over a finite range of scales using linear regression; see Fig.~\ref{fig:appBoxCounting}.
A well-defined scaling regime, visible as a straight line in this log-log plot, is a signature that the power law holds and the estimate of $D$ is meaningful.
Deviations at large scales reflect the finite extent of the attractor, whereas deviations at short scales arise once $\varepsilon$ becomes smaller than the typical point spacing in the cloud, so that individual points rather than the attractor geometry are being resolved.

From this analysis, we obtain a box-counting dimension $D \approx 1.5$. For comparison, the stationary measure $P_\infty$ has an integer-valued dimension equal to 2.
Interestingly, our numerical results indicate a weak dependence of the box-counting dimension on the particular realization of the attractor, \textit{i.e.}, on the noise history. The resulting distribution of box-counting dimensions exhibits a variance of order $0.1$.
Whether this is merely a finite-resolution effect that would disappear upon improving the numerical accuracy and extending the pullback evolution, or instead reflects the fact that the box-counting dimension is a genuinely random quantity with an extended distribution, remains an interesting question.

\subsection{Lyapunov exponent}
\begin{figure}
    \centering
    \includegraphics[width=0.51\linewidth]{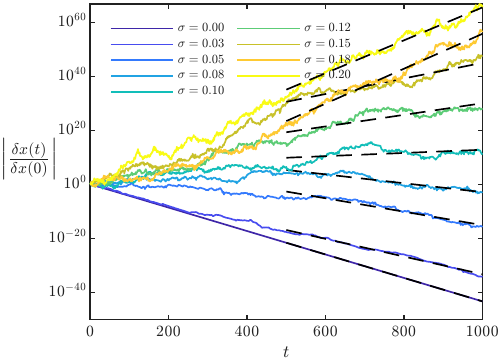}
    \includegraphics[width=0.48\linewidth]{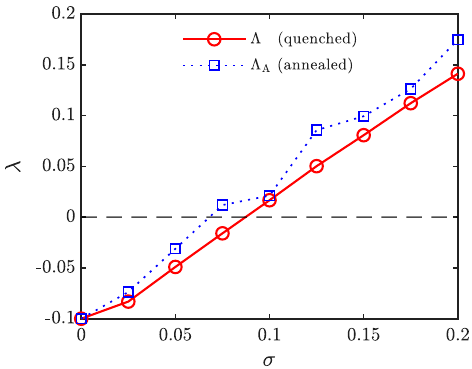}
    \caption{
    Butterfly effect associated with the stochastic dynamics in Eqs.~(\ref{app:eq:hopf_eq}).
    (a) $|\delta x(t)/\delta x(0)|$ as a function of $t$ for a single trajectory, for different noise strengths $\sigma$ given in the key (semilog plot). The dashed lines are the linear fits to determine $\Lambda_T$ in Eq.~(\ref{eq:Lambda_T}). 
    (b) Quenched and annealed Lyapunov exponents defined in Eqs.~(\ref{app:eq:def_Q}) and~(\ref{app:eq:def_A}) versus the noise strength $\sigma$. The data is computed from linear fits of the trajectories of $N_{\rm samples} = 10^3$ samples with different initial conditions and noise histories, as shown in the left.
    Parameters: $\mu = -0.1$, $\omega  = a =  1$, $b = 20$ ($\delta t = 10^{-4}$). 
    }
    \label{fig:appLyapunov}
\end{figure}
We now present the quantitative determination of the Lyapunov associated with the random attractor.

Let us first illustrate the butterfly effect in the case of a single trajectory, \textit{i.e.}, no averaging of any kind. In Fig.~\ref{fig:appLyapunov}~(a), we report the distance between trajectories generated by an infinitesimal perturbation of the initial conditions,
$\left|{\delta x(t)}/{\delta x(0)} \right|$, computed along single trajectories for different strengths of noise $\sigma$. 
Exponentially growing distances correspond to chaotic dynamics, while the others correspond to non-chaotic dynamics.

From each of these curves, we extract
\begin{align} \label{eq:Lambda_T}
    \Lambda_T := \frac{1}{T} \log \left|\frac{\delta x(T)}{\delta x(0)} \right|
\end{align}
by fitting $ \log\left|{\delta x(t)}/{\delta x(0)}\right|$ to a linear function in the interval $t\in [T/2,T]$.
Notably, for finite $T$, $\Lambda_T$ is a stochastic quantity that depends on the individual noise realization.
In principle, the limit $T\to\infty$ guaranties convergence towards a well-defined $\Lambda := \lim_{T\to\infty} \Lambda_T$. 
In practice, as the limit $T\to\infty$ is numerically challenging, we shall achieve this convergence by converging the solutions for a large but finite $T \gg  1/\Lambda_{T}$, and further averaging $\Lambda_T$ with respect to initial conditions and noise histories, which we denote as  $\langle \ldots \rangle$.
In Fig.~\ref{fig:appLyapunov}~(b), we plot the resulting quenched average
\begin{align} \label{app:eq:def_Q}
    \Lambda := \lim_{T\gg1/\Lambda} \langle \Lambda_T  \rangle,
\end{align}
where the average is performed over $N_{\rm samples} = 10^3$ samples. 
For completeness, we also plot the annealed version
\begin{align} \label{app:eq:def_A}
    \Lambda_{\rm A} := \lim_{T\gg1/\Lambda_{\rm A}}  \frac{1}{T} \log \Big\langle \left|\frac{\delta x(T)}{\delta x(0)} \right| \Big\rangle,
\end{align}
where the average over quantum fluctuations is performed inside the logarithm, making $\Lambda_{\rm A}$ the exponent associated with the exponential growth of the semiclassical out-of-time-order correlator (OTOC) rather than the Lyapunov exponent in the strict (quenched) sense.
Note that Jensen's inequality dictates $\Lambda_{\rm A} \geq \Lambda$.

\end{document}